\newcommand{\rmv}[1]{}
\newcommand{\conf}[1]{}

\newtheorem{theorem}{Theorem}
\newtheorem{lemma}{Lemma}
\newtheorem{proposition}[lemma]{Proposition}
\newtheorem{corollary}[lemma]{Corollary}
\newtheorem{definition}{Definition}

\def\proof{\noindent {\bf Proof. \hskip 0mm}}

\def\endproof{\hfill$ \Box$\vspace{0.5mm}}

\def\vd{{\boldsymbol d}} 
\def\vs{{\boldsymbol s}} 
\def\vv{{\boldsymbol v}} 

\def\vtype{{\boldsymbol \theta}} 

\def\valgo{{\boldsymbol A}}
\def\send{{\it send}}
\def\recv{{\it recv}}
\def\Msgs{{\sf Msgs}}

\def\smsgs{{\sf smsgs}}
\def\rmsgs{{\sf rmsgs}}
\def\mhist{{\sf mhist}}
\def\newepoch{{\sf NEWEPOCH}\xspace}

\def\MsgGraph{{\sf MsgGraph}\xspace}
\def\sent{{\small \sf sent}\xspace}
\def\notsent{{\mbox{\small \sf not-sent}}\xspace}
\def\uncertain{{\small \sf uncertain}\xspace}
\def\neverknown{{\small \mbox{\small \sf never-known}}\xspace}
\def\AlgUpdateMG{{\sc Alg-MsgGraph}\xspace}
\def\AlgDecideD{{\sc Alg-Dictator}\xspace}
\def\Alg2cheater{{\sc Alg-NewEpoch}\xspace}
\def\AlgConsistency{{\sc Alg-Consistency}\xspace}
\def\AlgWF{{\sc Alg-NewEpoch2}\xspace}
\def\AlgRand{{\sc Alg-RandNewEpoch2}\xspace}

\def\sender{{\it sender}}
\def\receiver{{\it receiver}}
\def\round{{\it round}}
\def\dictator{{\it dictator}}

\documentclass[11pt]{article}
\usepackage{times}
\usepackage{amssymb,amsmath}
\usepackage{graphics}
\usepackage{graphicx}
\usepackage{epsfig}
\usepackage{color}
\usepackage[vlined, ruled, linesnumbered]{algorithm2e}
\usepackage{fullpage}
\usepackage{xspace}
\usepackage{enumitem}

\begin{document}

\sloppy

\title{Distributed Consensus Resilient to \\ 
    Both Crash Failures and
	Strategic Manipulations}
\date{}
\author{Xiaohui Bei\\
Tsinghua University \\
beixiaohui@gmail.com
\and Wei Chen\\
Microsoft Research Asia \\
weic@microsoft.com
\and Jialin Zhang\\
University of Southern Califonia\\
zhangjl2002@gmail.com
}

\maketitle
\thispagestyle{empty}
\begin{abstract}
In this paper, we study distributed consensus in synchronous systems subject
	to both unexpected crash failures and strategic manipulations by
	rational agents in the system.
We adapt the concept of collusion-resistant Nash equilibrium to model
	protocols that are resilient
	to both crash failures and strategic manipulations of
	a group of colluding agents.
For a system with $n$ distributed agents, 
	we design a deterministic protocol that tolerates 
	$2$ colluding agents and a randomized protocol that
	tolerates $n-1$ colluding agents, and both tolerate any number of
	failures.
We also show that if colluders are allowed an extra communication round
	after each synchronous round, 
	there is no protocol that can tolerate even $2$ colluding agents and
	$1$ crash failure.

\end{abstract}

\setcounter{page}{1} 

\section{Introduction}

Consensus is a distributed task at the core of many distributed
	computing problems.
In consensus, each process proposes a value
	and eventually all processes need to agree on an irrevocable decision
	chosen from the set of proposed values.
Extensive studies have been conducted on consensus protocols tolerating
	various kinds of system failures from crash failures to malicious
	Byzantine failures (cf.~\cite{Lyn96,AW04}).


Besides unexpected system failures, users of distributed systems may alter
	their protocol components to achieve
	certain selfish goals, which we refer as {\em strategic manipulations}
	of distributed protocols.
The issue is more evident in systems spanning multiple administrative
	domains, such as peer-to-peer systems, mobile computing systems, and 
	federated cloud computing systems, where each computing entity has
	selfish incentives.
Combining unexpected Byzantine failures with strategic manipulations 
	in distributed protocol design
	have been studied in the context of secret sharing and multiparty
	computation~\cite{ADGH06,ADH08,LT06}, fault-tolerant
	replication~\cite{AACDMP05,CLNMAD08}, and gossip 
	protocols~\cite{LCWNRAD06}.
However, many important topics on incorporating selfish incentives with
	fault-tolerant distributed tasks left unexplored.
In particular, we are unaware of any work on incorporating selfish incentives
	with crash failures for distributed consensus.
Consensus protocols
	tolerating crash failures have been widely used in distributed systems,
	and thus it is natural to ask how to further tolerate strategic
	manipulations on top of crash failures.
Moreover, Byzantine consensus protocols cannot simply replace crash-resilient
	consensus protocols, because they only tolerate less number of failures,
	are more complex, and often require costly cryptographic
	schemes.
Byzantine consensus protocols cannot resist strategic manipulations either,
	because they only guarantees consensus and are not immune
	to manipulations that improve agents' utilities while satisfying consensus
	requirement.
Therefore, studying consensus protocols resilient to both crash failures
	and strategic manipulations are an important research topic of independent
	interest.


In this paper, we make the first attempt to
	tackle the problem of distributed consensus resilient to both crash 
	failures and strategic manipulations.
In particular, we study synchronous round-based consensus subject to
	both crash failures and manipulations by strategic {\em agents}
	(to differentiate from the mechanical processes) who have preference
	on consensus decisions.
As long as consensus is reached, an agent could manipulate his algorithm
	in arbitrary ways, such as faking
	the receipt of a message or pretending a crash failure, in order to
	reach a preferred decision value for him.
This models scenarios in which agents may want to gain access in mutual
	exclusion protocols or become the leader in leader election protocols,
	which are often implemented with a consensus component.

\begin{figure}[t]
 \centering
   \includegraphics[width=\textwidth]{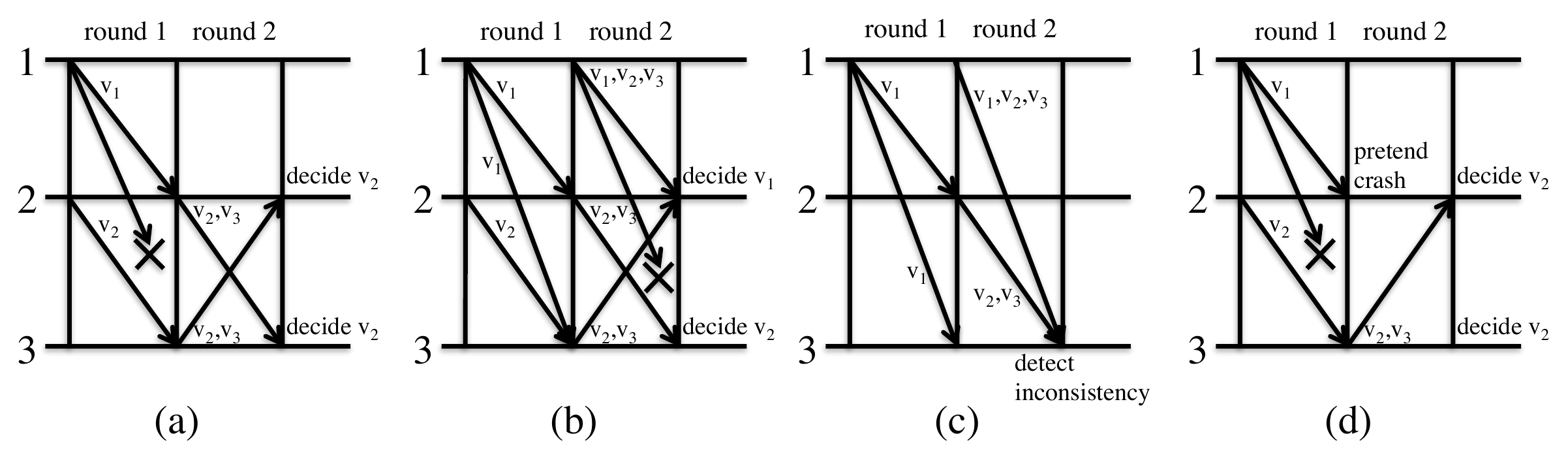}
 \caption{Motivating example on crash failures combined with
	strategic manipulations in a consensus protocol. Some unimportant
	messages are omitted for clarity.
	(a) A case of successful manipulation: agent $2$ manipulates his
	message to agent $3$ in round $2$ by not including
	$v_1$ in the message.
	(b) Consensus violation due to independent manipulations:
	agents $2$ and $3$ both manipulate their round-$2$ messages to each other
	by not including $v_1$.
	(c) Detecting inconsistency after manipulation: agent $2$ manipulates
	his round-$2$ message to agent $3$ as in case (a), but agent $3$ 
	detects this manipulation because it receives a message from
	agent $1$ in round $2$.
	(d) Successful manipulation without being detected: agent $2$
	pretends a crash in round $2$ to avoid sending a message to agent $3$.}
\label{fig:cases}
\end{figure}

Standard consensus protocols are easily manipulated, as shown by the following
	motivating example.
In a standard $n$-process
	synchronous consensus protocol tolerating $n-1$ crash 
	failures~\cite{Lyn96}, processes exchange all proposed values they
	received so far for $n-1$ rounds and at the end of round $n-1$ decide
	on the smallest value they received.
Now consider a simple system of three agents $\{1,2,3\}$, and agent $i$
	prefers value $v_i=i$ over other values and thus uses $v_i$ as his
	consensus proposal.
Agent $2$ can manipulate the above protocol in the following 
	way (Figure~\ref{fig:cases}(a)): 
	if in round $1$ agent $2$ receives the proposal $v_1$
	from $1$, he does not include $v_1$ in his message to agent $3$; then
	if in round $2$ agent $2$ does not receive a message from agent $1$ and 
	does not see $v_1$ in the message he receives from agent $3$, he knows that
	agent $1$ has crashed before sending his proposal to agent $3$.
In this case agent $2$ will decide his own proposal $v_2$, which is a better
	choice for him than $v_1$, the value he would decide 
	if he followed the protocol.
For agent $3$, he only sees agent $2$'s proposal $v_2$ and his own proposal
	$v_3$, 
	and since $v_2<v_3$, agent $3$ will follow the protocol
	and decide $v_2$, reaching consensus.
In all other cases, agent $2$ would follow the protocol, and it is easy to check
	that consensus is always reached.
Therefore, this standard synchronous consensus protocol is not resilient to
	strategic manipulations by a single selfish agent.

Strategic manipulations introduce uncertainty and instability to
	the system leading to unexpected system outcome, and thus should be
	prevented if possible in general.
In particular, consensus may be violated if more than one
	agents try to manipulate the protocol at the same time.
In the above example, agent $3$ may also conduct a manipulation
	symmetric to that of agent $2$, if he prefers $v_2$ over $v_1$.
Consider a run in which all agents are correct in round $1$ and in round
	$2$ agent $1$ crashes after sending a message to agent $2$ but before
	sending a message to agent $3$ (Figure~\ref{fig:cases}(b)).
Agents $2$ and $3$ independently
	want to manipulate the protocol, and thus in round
	$2$ they do not send $v_1$ to each other.
At the end of round $2$, according to their manipulation rule, agent
	$2$ would decide $v_1$ but agent $3$ would decide $v_2$, 
	violating consensus.
Therefore, resiliency to strategic manipulation is desirable to avoid
	such scenarios.

Designing a consensus protocol resilient to both crash failures and
	strategic manipulations is far from trivial, since one needs to cover
	all possible manipulation actions and their combinations.
Going back to the above motivating example, for the described cheating action
	of agent $2$, agent $3$ may detect inconsistency if he receives a message
	from agent $1$ in round $2$ but did not see $v_1$ from agent $2$'s message
	in round $2$ (Figure~\ref{fig:cases}(c)).
In this case, agent $3$ could execute a punishment strategy to hinder
	agent $2$ from taking the cheating action.
Even if certain cheating actions can be detected, one has to carefully go 
	through all possible cases and detect all of them.
More seriously, not all cheating actions can be detected.
For example, for agent $2$, instead of the above cheating action, he
	may pretend a crash in round $2$, not sending any message to agents
	$1$ and $3$ (Figure~\ref{fig:cases}(d)).
It is easy to see that agent $2$ can still benefit from this cheating action
	but the action cannot be detected by others.
Therefore, tolerating manipulations together with crash failures is
	a delicate task.
To make things more complicated, we further
	target at tolerating collusions of multiple
	agents.

In this paper, we 
	adapt collusion-resistant Nash equilibrium to model consensus protocols
	that resist both crash failures and strategic manipulations.
Roughly speaking, we say that a group of colluders can manipulate 
	a consensus protocol if they can change their protocol execution
	such that the deviation still guarantees consensus in all possible
	crash scenarios, and in one crash scenario one of the colluders
	benefits (i.e. obtains a better consensus decision).
We say that a consensus protocol is {\em $(c,f)$-resilient} if it 
	solves consensus and no group of colluders of size at most $c$ 
	can manipulate the protocol, in a synchronous system
	with at most $f$ crash failures.
Our contributions in this paper include:



\begin{itemize}
\setlength{\itemsep}{-1mm}

\item
We propose the solution concept of $(c,f)$-resilient consensus protocol
	to incorporate selfish behaviors into crash-prone systems
	in distributed protocol design, and connecting it with the theory
	of social choices.

\item
We provide a deterministic $(2,f)$-resilient consensus protocol and
	a randomized
	$(n-1,f)$-resilient consensus protocol for any $f\le n-1$.
Both protocols
	are polynomially bounded in round complexity, message complexity,
	and local computation steps, and neither of them relies on any cryptography 
	or computational hardness assumptions.


\item
Moreover, we show that if colluders have an extra round of
	communication after each synchronous round ends, 
	then no $(2,f)$-resilient consensus protocol (even randomized) exists
	for any $f\ge 1$. 
Thus extra communication among colluders are very powerful for strategic
	manipulations.
\end{itemize}

Our study demonstrates both the feasibility and difficulty in incorporating
	selfish incentives with crash failures, and provides several techniques
	that could be used to defense against strategic manipulations in other
	similar situations.
We hope that our study could lead to more work on incentive-compatible
	and fault-tolerant distributed protocols.

\subsection{Related work}

Both fault-tolerant distributed computing and game theory 
	address distributed entities that may experience
	abnormal behaviors, whether being system failures or strategic
	manipulations.
The combination of the two, however, is not yet widely explored.
Several existing studies address the combination of
	Byzantine failures with strategic 
	manipulations~\cite{AACDMP05,LCWNRAD06,CLNMAD08,ADGH06,ADH08,LT06}
	in distributed protocol design.

The BAR (standing for Byzantine, Altruistic, and Rational) fault tolerance
	framework, proposed in a series of 
	work~\cite{AACDMP05,LCWNRAD06,CLNMAD08}, incorporates rational behavior
	into Byzantine fault-tolerant distributed protocols for state machine
	replication, gossip, terminal reliable broadcast, and backup services
	built on these protocols.
They also consider agent utilities that prefer certain proposed values,
	but their solution concept is weaker than ours in two aspects.
First, a rational agent would only deviate from the given protocol
	when he could benefit no matter who are the Byzantine agents and
	what the Byzantine agents would behave, while in our model
	a rational agent could deviate as long as it guarantees consensus and
	receives benefit in {\em one possible} crash scenarios.
Second, they do not consider collusions among rational players.	
As a result, their protocol does not apply to our setting.

In~\cite{ADGH06,ADH08}, Abraham et al. study rational secret sharing and
	multiparty computation that could 
	tolerate both Byzantine failures and a colluding group of rational agents,
They provide matching upper and lower bounds on implementing a trusted
	mediator resilient to $k$ colluders and $t$ Byzantine agents
	with distributed protocols also resilient to $k$ colluders and
	$t$ Byzantine agents.
Their solution concept of $(k,t)$-robust equilibrium is stronger than ours
	in that (a) a rational agent could deviate as long as he benefits
	in one possible Byzantine failure scenario, even if the deviation may
	lead to drastic damages (e.g. violation of consensus) in other situations;
	and (b) Byzantine failures can only improve the utility of non-Byzantine
	agents.
Even though their solution concept is stronger, their positive results
	do not apply to our consensus setting, 
	because they need the condition that a $(k,t)$-robust equilibrium
	with a trusted mediator exists, but in our consensus setting such
	equilibria do not exist with any $t\ge 1$
	(See discussion in the Section~\ref{sec:justify}).


In~\cite{LT06}, Lysyanskaya and Triandopoulos also consider both 
	rational and Byzantine behaviors in the context of multiparty 
	computation.
They study a class of rational behaviors in which a rational agent gets
	increased utility if his computation gets close to the target value or
	other agents' computation gets further away from the target value.
In our case, an agent would not get an increased utility simply because
	other agents are getting worse.

There are a few other studies that address
	both Byzantine failures and strategic manipulations.
In~\cite{MSW06}, Moscibroda et al. study a virus inoculation game in which
	some players are Byzantine, and they define price of malice as a way
	to measure the impact of Byzantine players to system efficiency.
In~\cite{BHLR08}, Blum et al. propose the use of regret minimization instead
	of Nash equilibrium in the study of price of anarchy, and define the
	price of total anarchy, which is robust against Byzantine players who
	do not play best responses.
These studies, however, focus on introducing Byzantine behavior to the
	strategic game format, while our work together with~\cite{AACDMP05,ADGH06}
	focus on introducing rational behavior into fault-tolerant
	distributed computing protocol design.

A number of other works address strategic behavior in distributed protocol
	design, but they do not further consider unexpected system failures.
Distributed algorithmic mechanism design (DAMD) is a framework proposed
	in~\cite{FPS01,FS02} and later refined in~\cite{PS04,SP04,FPSS05},
	which incorporates rational behaviors in the design of distributed
	protocols such as multicast cost sharing and internet routing.
Several studies address secret sharing and multiparty computation among
	rational agents~\cite{HT04,IML05,GK06}.
A concurrent work by Abraham et al.~\cite{ADH11} studies leader election
	with rational agents. 
They consider both synchronous and asynchronous networks with fully connected
	or ring network topologies, where agents prefer to be elected as the
	leader.
Since they do not consider failures, their focus is on developing 
	strategy-proof random selections so that every agent has a fair chance
	to be the leader.


\paragraph{Paper organization.}
In Section~\ref{sec:model}, we describe our model for consensus protocols
	that resist both crash failures and strategic manipulations, and
	also connect our model with a classic result on social choice functions.
Section~\ref{sec:cons} contains our main algorithm results
	on $(c,f)$-resilient consensus protocols.
We start by presenting a deterministic $(2,f)$-resilient protocol
	for $f \le n-2$ in Section~\ref{sec:2cheater}, and then extend it
	to a deterministic $(2,f)$-resilient protocol for
	$f \le n-1$ in Section~\ref{sec:WF} and a randomized $(n-1,f)$-resilient
	protocol for $f\le n-1$ in Section~\ref{sec:rand}.
We discuss protocol complexity and summarize techniques used in our
	protocols for resisting
	strategic manipulations in Section~\ref{sec:tech}.
Section~\ref{sec:imp} shows the impossibility result on tolerating
	colluders with extra communication rounds.
We conclude the paper and discuss future directions in
	Section~\ref{sec:conclude}.

\section{Modeling consensus resilient to strategic manipulations and 
	crash failures}
\label{sec:model}

\subsection{System model}

We consider a distributed 
	system with $n$ agents $\Pi = [n] = \{1,2,\ldots, n\}$. 
Agents proceed in synchronous rounds, with round $1$, $2$, and so on. In each round, agents start by sending messages to a selected set of other agents. After sending out messages, agents will receive messages sent to them in the same round, and then
	based on received messages update their local states. 
Agents will not receive messages sent in earlier rounds. The channels are reliable, that is, if a message is sent by agent $i$ to agent $j$  in round $r$,  and neither of them crash in this round, then agent $j$ will receive this message from agent $i$ at the end of round $r$. 
Formally, the synchronous round system provides two interface functions
	$\send()$ and $\recv()$.
The invocations of $\send()$ and $\recv()$ have to be well-formed, more
	specifically, the invocation sequence on an agent has to start with
	a $\send()$, and then alternates between $\recv()$ and $\send()$.
The $r$-th pair of $\send()$ and $\recv()$ is for synchronous round $r$.
Let $\cal M$ be the set of all possible messages sent in the system.
Let $\Msgs$ be an array type $[n]\rightarrow {\cal M}\cup \{\bot\}$.
Interface $\send()$ takes one parameter $\smsgs \in \Msgs$, such that
	if $\smsgs[j]\in {\cal M}$, 
	$\smsgs[j]$ is the message the calling agent wants to send to agent
	$j$ in this round, and if $\smsgs[j]=\bot$, it means the calling agent
	does not want to send any message to $j$ in this round.
Interface $\recv()$ returns a result $\rmsgs \in \Msgs$, such that
	if $\rmsgs[j]\in {\cal M}$, $\rmsgs[j]$ is the message the caller
	receives from agent $j$ in this round, and if $\rmsgs[j]=\bot$, it means
	that the caller does not receive any message from $j$ in this round.

The round model is fixed by the system and cannot be manipulated by
the agents. For example, an agent cannot wait to receive messages of
round $r$ and then send out his own message of round $r$ that may
depend on the messages received. More precisely, an agent can only
manipulate $\smsgs$ in his $\send(\smsgs)$, and not others, including
the alternating sequence of $\send()$'s and $\recv()$'s.

Agents may fail by crashing, and when an agent $i$ fails in a round $r$, 
	it may fail to send its round-$r$ messages to a subset of agents, and
	it will stop executing any actions in round $r+1$ or higher.
Formally, agent failures are characterized by a failure pattern $F$,
	which is 
	defined as a subset of $\{(i,j,r)\ |\ i,j\in \Pi, r= 1,2,3,\ldots\}$,
	and satisfies the constraint that if $(i,j,r) \not\in F$, then
	for all $r'> r$, for all $j'\in \Pi$, $(i,j',r')\not\in F$.
Thus $(i,j,r)\in F$ means that $i$'s message to $j$ in round $r$ would be
	successful should $i$ send a message to $j$ in $r$.
For some agent $i$, if for all $j\in \Pi$ and all $r\ge 1$, $(i,j,r)\in F$, 
	then we say that agent $i$ is {\em non-faulty (or correct)}; otherwise,
	we say that agent $i$ is {\em faulty}.
We say that $i$ {\em crashes in round $r$} if
	$r$ is the smallest round such that there is some $j$ with
	$(i,j,r)\not\in F$, and $i$ {\em is alive in round $r$} if
	for all $j\in \Pi$, 
	$(i,j,r)\in F$.\footnote{We consider the case that $i$ crashes
	after successfully sending out all round $r$ messages but before 
	receiving round $r$ messages or updating local states as the same as
	$i$ crashes at the beginning of round $r+1$ before sending out any
	messages of round $r+1$, since no other agent can distinguish these
	two cases.}
In the paper, we use $f$ to represent the number of possible faulty agents in 
	an execution.

Let $V$ be a finite set of possible consensus proposal values. 
The {\em private type (or type)} $\theta_i$ of agent $i$ is 
	his preference on the set of proposals $V$. 
That is, it is a total order $\prec_i$ on $V$, and $u\prec_i v$ means $i$ prefers $v$ over $u$. 
Let $\vtype = (\theta_1, \theta_2, \ldots, \theta_n)$ be the
   vector of private types.

The {\em message history} of agent $i$ at the end of round $r$
	is an array $\mhist_i[1..r]$, where $\mhist_i[k]\in \Msgs$ denotes the
	messages $i$ receives in round $k$, for $k=1,2,\ldots, r$.
The {\em local state} of agent $i$ at the end of round $r$, i.e.,
	before $i$ invokes its \mbox{$(r+1)$-th} $\send()$, includes
	(a) the current round number $r$, 
	(b) the message history $\mhist_i[1..r]$ of $i$, and
	(c) his private type $\theta_i$.
As a convention, when $r=0$, the message history $\mhist_i[1..r]$ is empty,
	which represents the initial state at the beginning of the first round.

A {\em deterministic algorithm} $A_i$ of 
	agent $i$ is a function from his local state at
	the end of each round to the messages it is going to send in the 
	next round, i.e.,
	$A_i(r,\mhist_i[1..r],\theta_i)$ is the parameter in agent $i$'s
	$(r+1)$-th invocation of $\send()$, for $r=0,1,2,\ldots$.
Let $\valgo$ denote the collection of algorithms $A_1,A_2, \ldots, A_n$,
	which we also refer as a {\em protocol}.
Given a failure pattern $F$, a private type vector $\vtype$, 
	and a deterministic protocol $\valgo$, the full execution of the system is determined, 
	and we call
   the execution a {\em run}.
A run describes exactly on every round what are the messages sent and received
   by every agent.
Let $R(F, \vtype, \valgo)$ denote this run.
We will introduce randomized protocols later in Section~\ref{sec:rand}.


\subsection{Rational consensus with strategic manipulations}

For the consensus task, each agent $i$ has an output variable $d_i$ with
   initial value $\bot\not\in V$.
In every run, $d_i$ is changed at most once by $i$ to a new value in 
	$V\cup \{\top\}$,
   which is called the consensus decision of $i$.
The special symbol $\top\not\in V$ is not among any proposal values and is used 
	as a punishment strategy by agents, which will be clear when we introduce
	our consensus protocol in Section~\ref{sec:cons}.
We use $\vd(F, \vtype, \valgo)$ to denote the decision vector when
   the run $R(F, \vtype, \valgo)$ completes and
   $d_i(F, \vtype, \valgo)$ to denote the decision
   of $i$ in the run.
Note that some decisions may still be $\bot$, either because the agent
   has crashed, or the agent chooses not to decide.
To incorporate decision output to the model, we slightly modify the
	definition of algorithm $A_i$ of agent $i$, such that 
	the function value $A_i(r,\mhist_i[1..r],\theta_i)$ also includes
	the decision value $d$, which could be $\bot$ meaning that $i$ 
	does not decide at the end of round $r$, or an actual value meaning that
	$i$ decides on this value at the end of round $r$.

To align with the standard terminology in game theory, 
	we define {\em strategies} of
	agent $i$ as follows.
Note that algorithm $A_i$ of $i$ is a function
	$A_i:(r,\mhist_i[1..r],\theta_i) \mapsto (\smsgs,d)$, we define
	a {\em strategy} $s_i$ of agent $i$ to be essentially the same 
	as $A_i$, but re-arrange it to be a function from the private type 
	$\theta_i$, that is 
	$s_i : \theta_i \mapsto ( (r,\mhist_i[1..r] )\mapsto (\smsgs,d))$, 
	such that $s_i(\theta_i)(r,\mhist_i[1..r] ) =
	A_i(r,\mhist_i[1..r],\theta_i)$.
Let $\vs = (s_1, s_2, \ldots, s_n)$ denote a {\em strategy profile}, and
	let $\vs(\vtype)$ denote 
	$(s_1(\theta_1), s_2(\theta_2), \ldots, s_n(\theta_n))$.

We focus on distributed consensus in this paper, which requires all
	agents eventually decide on the same value.
We specify the consensus task
	in terms of the following legal strategy profile.

\begin{definition}[legal strategy profile w.r.t consensus, a.k.a. 
	consensus protocol]
A strategy profile $\vs$ is {\em legal with respect to consensus} if for
   any failure pattern $F$ and any private type vector $\vtype$,
   the resulting run $R(F, \vtype, \vs)$ always satisfies the following
   properties:
\begin{itemize}
\setlength{\itemsep}{-1ex}
\item Termination: every correct agent eventually decides in the run.

\item Uniform Agreement: no two agents (correct or not) decide differently.

\item Validity: if some agent $i$ decides $v\in V$, then 
	$v$ must be the most preferred value of some agent $j$ (according
	to $\theta_j$).


\end{itemize}

\end{definition}

Note that a strategy profile is essentially a protocol (collection of
	agents' algorithms), and thus we will use the terms strategy profile and
	protocol interchangeably.
In this paper, we consider the {\em utility} $u_i$ of agent $i$ to be
	only dependent on the decision vector of a run, and is consistent
	with $i$'s preference specified in his private type $\theta_i$.
Formally, given a run $R(F, \vtype, \vs)$ under failure pattern $F$, private
	type vector $\vtype$ and strategy profile $\vs$, we define the utility
	of $i$ in this run, $u_i(R(F, \vtype, \vs), \theta_i)$, to be:
	(a) if $i$ crashes according to $F$ in the run, then 
	$u_i(R(F, \vtype, \vs), \theta_i)=0$; 
	(b) if the run satisfies
	consensus properties (Termination, Uniform Agreement and Validity) with
	the decision value $d$, then 
	$u_i(R(F, \vtype, \vs), \theta_i)$ is a positive value such that
	the more $i$ prefers $d$, the higher the utility value; and
	(c) if the run violates at least one of the consensus properties,
	$u_i(R(F, \vtype, \vs), \theta_i)= -\infty$.
This utility function indicates that agent $i$ is neutral if he crashes
	in the run, he has preference on the decision value when consensus
	is satisfied, and it would be a disaster for him if consensus is violated.

In this paper, we allow a group of colluding agents to manipulate the
	protocol together in order to benefit one of the colluders, which
	we formalize below.
Let $C\subseteq \Pi$ be the set of colluding agents.
We allow colluders to know the private types of all colluders in advance.
To model this, we say that a colluder $i$'s strategy $s_i$ is a function
	$s_i : \theta_C \mapsto ( (r,\mhist_i[1..r] )\mapsto (\smsgs,d))$, 
	where $\theta_C$ is a sub-vector containing all entries for all $j\in C$.
We use $\vs'_C$ to denote {\em colluders' strategy profile}, 
	which contains entries for $i\in C$ where each $\vs'_i$ is a function
	from $\theta_C$ as defined above.
We denote
	$(\vs_{-C}, \vs'_C)$ as a new vector obtained from strategic
	profile $\vs$ by
	replacing all entries of $\vs$ for $i\in C$ with the corresponding
	entries in $\vs'_C$.

\begin{definition}[group strategic manipulation]
\label{def:groupcheat3}
For a legal strategy profile $\vs= (s_1, s_2, \ldots, s_n)$,
	a group of agents $C$ can strategically manipulate 
	profile $\vs$ if there exists a colluders' strategy profile
	$\vs'_C$ for agents of $C$, such that
	(a) $(\vs_{-C}, \vs'_C)$ is still a legal strategy profile, and
	(b) there exists a failure pattern $F$ and
    a private type vector $\vtype$ in which all agents in 
	$C$ have the same most preferred proposal, 
    such that the utility of some $i\in C$ in $(\vs_{-C}, \vs'_C)$ is
    better off: $
    u_i(R(F, \vtype, (\vs_{-C}, \vs'_C)),
    \theta_i) > u_i(R(F, \vtype, \vs), \theta_i)$.
\end{definition}

We also refer to colluding agents as {\em cheaters},
	and agents who follow the protocol as {\em honest
	agents}.
We are now ready to introduce our central solution concept.

\begin{definition}[$(c,f)$-resilient equilibrium,
	or $(c,f)$-resilient consensus protocol]
\label{def:colludenash1}
A {\em $(c,f)$-resilient equilibrium, or 
	$(c,f)$-resilient consensus protocol}
	is a legal strategy profile
   $\vs $ for consensus, such that no group of agents of
	size at most $c$ can strategically
   manipulate profile $\vs$, in a system with at most $f$ crash failures.
\end{definition}



Note that when $c=0$, $(c,f)$-resilient equilibria are simply classic
	synchronous consensus protocols tolerating $f$ crash failures.
When $c=1$, if we remove the legal strategy profile requirement
	(condition (a) in Definition~\ref{def:groupcheat3}), our solution concept
	would match the ex-post Nash equilibrium concept.
Our legal strategy profile requirement makes our solution concept unique,
	and we will justify its inclusion shortly.
Before providing justifications to our solution concept, we will first
	make a connection of our solution concept with the theory of
	social choice functions and state an important result, which will be
	used in our justifications.

\subsection{Dictatorship in ex-post Nash equilibrium}

Given a legal strategy profile $\vs$ and a failure pattern $F$, we say
	that an agent $i$ is a {\em dictator} 
	of $\vs$ under $F$ if for any private type
	vector $\vtype$, the decision in run $R(F,\vtype, \vs)$ is always
	the most preferred value of $i$.
We now connect our solution concept with social choice functions and
	establish the result that for any $c\ge 1$ and $f\ge 0$, 
	any $(c,f)$-resilient equilibrium under
	any failure pattern must have a dictator.
A {\em social choice function} $f$ (in our context) is a function from
	a private type vector $\vtype$ to a value in $V$.
A social choice function $f$ is {\em incentive compatible} if
	there does not exist an agent $i$, a private type vector $\vtype$,
	two proposal values $a,b\in V$, such that (a) $f(\vtype)=b$, 
	(b) $i$ prefers $a$ over $b$ in $\theta_i$, and 
	(c) $i$ could find another private type $\theta_i'$
	so that $f((\vtype_{-i}, \theta_i')) = a$
	(c.f Chapter 9 of~\cite{NRTV07}).
We say that a social choice function $f$ is a {\em dictatorship}
	if there exists an agent $i$ such that for all private type
	vector $\vtype$, $f(\vtype)$ is always the most preferred value
	in $\theta_i$.
The following is the famous Gibbard-Satterthwaite
    Theorem on incentive compatible social choice functions, which is also
	a version of Arrow's Impossibility Theorem on social welfare functions.
\begin{proposition}[Gibbard-Satterthwaite Theorem~\cite{Gib73,Sat75}]
If $f$ is an incentive compatible social choice function onto $V$ and
	$|V|\ge 3$, then $f$ is a dictatorship.
\end{proposition}

\def\thmdictator{
For any $c\ge 1$, any failure pattern $F$ with at most $f$ crash failures,
	if $\vs$ is $(c,f)$-resilient equilibrium and $|V|\ge 3$, 
	then there always exists a dictator of $\vs$ under $F$.
}

\begin{theorem}\label{thm:dictator} 
\thmdictator
\end{theorem}
\begin{proof}
Consider any $(c,f)$-resilient equilibrium $\vs$ and a failure pattern $F$
	with at most $f$ failures.
The strategy profile $\vs$ under $F$ can be viewed as a social choice
	function $f_{\vs,F}$ from private type vector $\vtype$ to the decision value
	in the run $R(F,\vtype, \vs)$. 
Since $\vs$ is an $(c,f)$-resilient equilibrium with $c\ge 1$, 
	we know that its
	corresponding social choice function $f_{\vs,F}$ is incentive
	compatible.
In fact, if it is not the case, then we can find an agent $i$, 
	a private type vector $\vtype$, a private type $\theta_i'$ of $i$, 
	such that $f_{\vs,F}(\vtype)=b$, $f_{\vs,F}((\vtype_{-i}, \theta_i')) = a$, and
	$i$ prefers $a$ over $b$ in $\theta_i$.
If so, agent $i$ can choose an alternative strategy $s_i'$
   such that $s_i'(\theta_i) = s_i(\theta_i')$ and for all other
	$\theta_i'' \ne \theta_i$, $s_i'(\theta_i'') = s_i(\theta_i'')$.
Essentially when $i$'s type is $\theta_i$, he just pretends his type
	is $\theta_i'$.
Since $\vs$ is a legal strategy profile, 
	$(\vs_{-i},s_i')$ is also a legal strategy profile, because
	$i$ only changes his type but nothing else.
However, agent $i$ could choose $s_i'$ so that he will benefit
	under private type vector $\vtype$ and failure pattern $F$, 
	contradicting to the fact
	that $\vs$ is a $(c,f)$-resilient equilibrium.

Moreover, since all runs $R(F,\vtype, \vs)$ satisfy Validity of consensus,
	we know that any proposal in $V$ could be a possible decision value.
That is $f_{\vs,F}$ is onto $V$.
Thus by the Gibbard-Satterthwaite
    Theorem, $f_{\vs,F}$ must be a dictatorship, i.e., there always exists
	a dictator of $\vs$ under $F$.
\end{proof}

Henceforth, we assume $|V| \ge 3$, which means
	from the above theorem that any $(c,f)$-resilient 
	consensus protocol with $c\ge 1$ must be a dictatorship for any
	failure pattern $F$.

\subsection{Explanation and justification of the solution concept}
\label{sec:justify}

We are now ready to provide some explanations and justifications to 
	our solution concept.

\newcounter{remcnt}

\addtocounter{remcnt}{1}
\paragraph{Remark \arabic{remcnt} (On possible cheating behaviors).}
We allow an agent to cheat not only
	by faking a different private type, but also by modifying his 
	entire algorithm, such as sending messages he is not supposed to send,
	pretending the receipt of messages he does not actually receive, or
	pretending to have a crash failure, etc.
An agent has all the freedom to change his algorithm (i.e. strategy) as long
	as he ensures that the changed algorithm together with
	other agents' algorithm still guarantees consensus, for all possible
	failure patterns and private type vectors.

\addtocounter{remcnt}{1}
\paragraph{Remark \arabic{remcnt}  (On legal strategy profile requirement).}
The requirement that the strategy profile after manipulation
	is still legal 	(condition (a) in Definition~\ref{def:groupcheat3})
	distinguishes our
	solution concept from the existing treatments combining Byzantine
	failures with strategy 
	manipulations~(\cite{AACDMP05,CLNMAD08,ADGH06,ADH08,LT06}.
In our setting, colluders have to ensure that in all possible failure patterns
	consensus is reached after the deviation, which
	means that rational agents are risk-averse in terms of 
	reaching consensus.
However, conditioned on that consensus is always ensured, rational agents
	would deviate from the protocol as long as there exists some failure
	pattern and some type vectors of agents
	in which some of them benefits, which means they are risk-taking
	under the condition that consensus is guaranteed.

Our solution concept captures a natural situation where the game outcomes
	could be normal or disastrous, and agents tend to be risk-averse
	in avoiding disasters but risk-taking when they are sure the disaster
	would not happen by their manipulations.
In the case of consensus, violation of consensus could be 
	disastrous, for example,
	it may lead to inconsistent copies in state machine replications, or
	concurrent access of critical resources in mutual exclusion protocols, which
	may bring down the entire system. 
Therefore, agents are only willing to manipulate the protocol when they
	are sure consensus would not be violated.

Our solution concept can be further explained if we consider that
	agents may have partial knowledge about the system, e.g.,
	agents may know the probability distribution of 
	failure patterns and type vectors of agents.
In this case, if there is a non-zero probability that consensus is violated,
	it will give the agent a $-\infty$ payoff since the utility of
	the disastrous outcome
	of violating consensus is $-\infty$ , and thus agents will always
	avoid consensus violation.
On the other hand, if with probability one
	consensus is satisfied, then as long as the gain of the agents under some
	failure patterns and type vectors outrun the loss under other cases,
	the agents would deviate.
By using our solution concept, we model the above situation without
	the complication of modeling probabilistic events in the system.

Moreover, in the consensus setting
	our requirement on legal strategy profile is also necessary.
This is easy to see when $c+f=n$: If we do not require that the deviated
	strategy profile to be legal, then the $c$ colluders can simply decide
	on their most preferred value without any communication, hoping that
	the rest $n-c=f$ agents are all crashed, which means no $(c,f)$-resilient
	consensus protocol exists.
The following proposition further shows that this is the case for
	all $c,f\ge 1$.

\begin{proposition} \label{prop:legal}

If we remove
	the legal strategy profile requirement specified as 
	condition (a) in Definition~\ref{def:groupcheat3}, then
	no $(c,f)$-resilient
	consensus protocol exists for any $c,f\ge 1$.
\end{proposition}
\begin{proof}
For a contradiction, let $\vs$ be 
	such a $(c,f)$-resilient consensus protocol.
Note that $\vs$ is also an $(c,f)$-resilient consensus protocol
	with the legal strategy profile requirement, and thus
	Theorem~\ref{thm:dictator} still applies to $\vs$.
Let ${\cal R}(F)$ be the set of runs $R(F,\vtype, \vs)$ under failure
	pattern $F$, with all possible type vectors $\vtype$.
By the Termination property of consensus, in all these runs all non-faulty
	agents decide.
Let $r_F$ be the largest round number at which some agent decide, among
	all runs in ${\cal R}(F)$. 
Since $V$ is finite, the number of possible type vectors $\vtype$ is also
	finite, and thus set ${\cal R}(F)$ is finite and 
	$r_F$ is a finite number.

Consider the failure-free pattern $F_0$.
By Theorem~\ref{thm:dictator}, there is a dictator $d$ of $\vs$ under
	failure pattern $F_0$.
Let $F_1$ be a failure pattern in which $d$ crashes at the beginning
	of round $r_{F_0}+1$, and all other agents are correct.
Since all agents have decided by the end of round $r_{F_0}$ in all runs
	with failure pattern $F_0$, crashing $d$ at the beginning of
	round $r_{F_0}+1$ does not change any decision value.
Thus $d$ is also the dictator of $\vs$ under $F_0$.
Let $F^*$ be another failure pattern in which $d$ crashes at the beginning
	without sending any messages. 
Let $d^*$ be the dictator of $\vs$ under $F^*$.
It is clear that $d^*$ is different from $d$, since in $F^*$ $d$ has no 
	chance to send any messages, no other agents would know the most
	preferred value of $d$.

Now consider a graph with all possible failure patterns as vertices,
	and two vertices $F$ and $F'$ have an edge if one has one more
	entry $(i,j,r)$ than the other.
On this graph, we can find a finite path from $F_1$ to $F^*$, since we
	can remove the entry $(d,j,r)$ one by one from $F_1$, starting from
	round $r_{F_0}$.
Since the two ends of this path has different dictators $d$ and $d^*$,
	respectively, along the path from $F_1$ to $F^*$, we can find the
	first edge from $F_a$ to $F_b$ such that the dictator changes to be
	another dictator $d''$ different from $d$.
Let $(d,j,r)$ be the additional entry that $F_a$ has comparing to $F_b$.

We argue that agent $j$ can manipulate protocol $\vs$, if we do not
	have the legal strategy profile requirement on the manipulation.
Agent $j$'s manipulation is as follows.
At the end of round $j$, when $j$ receives a message from $d$, he
	simply pretends that he does not receive any message from $d$ in round
	$r$ and acts accordingly in the later rounds.
Agent $j$ will gain the benefit if the run has failure pattern $F_a$ and
	he prefers the proposal of $d''$ over that of $d$, 
	since with his deviation 
	all agents would behave as if they are in failure pattern $F_b$ and
	decide on the proposal of $d''$.
Therefore, $\vs$ is not a $(c,f)$-resilient consensus protocol for any
	$c,f\ge 1$.
\end{proof}

With the legal strategy profile, however, the manipulation of $j$ stated
	in the above proof would be easily handled with, because
	it is possible that $d$ does not crash in round $r$ and continues
	sending messages in round $r+1$, so that other agents would immediately
	detect that $j$ has manipulated the protocol, and execute certain
	punishment strategy on $j$.
Therefore, the legal strategy profile requirement is both reasonable
	and necessary in our setting.

\addtocounter{remcnt}{1}
\paragraph{Remark \arabic{remcnt} (on all colluders having the same
	most preferred proposal).}
In Definition~\ref{def:groupcheat3}, we require that all colluders in 
	$C$ have the same most preferred proposal. 
Without this requirement, no protocol can escape a trivial cheating
	scenario in which a colluding dictator simply uses his colluding
	pattern's most preferred value instead of his own, 
	as detailed by the proposition below. 
Moreover, it is also reasonable to assume that colluders having the same most
	preferred value, since they usually share some common goal,
	which is what brings them together in the first place.

\begin{proposition}
If we do not require that all colluders have the same most preferred
	value, then there is no $(c,f)$-resilient consensus protocol
	for any $c\ge 2$.
Moreover, even if we require that no colluder is worse off
	in condition (b) of Definition~\ref{def:groupcheat3}, 
	there is no $(c,f)$-resilient consensus protocol for any
	$c\ge 2$ and $f\ge 1$.
\end{proposition}
\begin{proof}
Suppose, for a contradiction, that there is a 
	$(c,f)$-resilient consensus protocol $\vs$, when
	we  do not require that all colluders have the same most preferred
	value.
Consider the failure-free failure pattern $F$.
Note that Theorem~\ref{thm:dictator} only concerns non-colluding
	deviation, so it still applies in this case to $\vs$.
Let $d$ be the dictator of $\vs$ under failure pattern $F$.
Let $i\neq d$ be another agent colluding with $d$, and $d$ and $i$
	have different most preferred values $v_d$ and $v_i$, respectively.
Then $d$ could simply pretend that his most preferred value is $v_i$.
Since $d$ is the dictator, the consensus decision would be $v_i$, 
	which is a strictly better result for $i$.
Since only the type of $d$ is changed, the resulting strategy profile must
	also be legal.
Therefore, we can a case of $\{d,i\}$ colludes and they manipulate the
	protocol so that $i$ benefits from the manipulation.
This contradicts to the assumption that $\vs$ is $(c,f)$-resilient.

In the above manipulation, $d$ is worse-off after the manipulation.
However, if $f\ge 1$, we can let $d$ crashes at the end after consensus
	decision is made, which means in the new failure pattern $d$
	is still the dictator.
In this failure	pattern, $d$'s utility remain the same since he crashes,
	and $i$'s utility is better-off, so we have a case that
	no colluder is worse-off if we allow at least one crash failure.
\end{proof}

\addtocounter{remcnt}{1}
\paragraph{Remark \arabic{remcnt}  (on one colluder benefiting from
	the deviation).}
In Definition~\ref{def:groupcheat3}, we allow some of the colluders to be
	worse off as long as one colluder is better off.
In fact, in our setting it is equivalent to requiring all non-faulty
	colluders are better off (the faulty agents always have utility
	$0$ by definition, so they are not worse off), 
	as shown by the following proposition.

\begin{proposition}
Assume that we change
	the condition (b) of Theorem~\ref{thm:dictator} such that we require
	all non-faulty colluders in $C$ have 
	to be better off after the manipulation.
Let $\vs$ is a $(c,f)$-resilient consensus protocol under this new
	definition for any $c \ge 1$.
Then $\vs$ is also a $(c,f)$-resilient consensus protocol under
	the old definition.
\end{proposition}
\begin{proof}
Note that Theorem~\ref{thm:dictator} only concerns the deviation of a
	single agent, thus it still applies to $\vs$ under the new definition.
Suppose, for a contradiction, that there exists a set $C$ of colluders, such
	that they can manipulate $\vs$ to be another legal strategy profile
	$\vs'=(\vs'_C, \vs_{-C})$ under the old definition.
Then there must exist a failure pattern $F$, a type vector $\vtype$, 
	and an agent $i\in C$, such that $i$ decides $v_1$ in the run
	of $\vs$ with $F$ and $\vtype$, while $i$ decides $v_2$ in the
	run of $\vs'$ with $F$ and $\vtype$, and $i$ prefers $v_2$ over $v_1$
	in $\theta_i$.
Since $\vs$ is a $(c,f)$-resilient equilibrium under the new definition,
	by Theorem~\ref{thm:dictator} there exists a dictator $d$
	of $\vs$ under failure pattern $F$.

If $d\in C$, then in the run of $\vs$ under failure pattern $F$ and 
	type $\vtype$, agents decide on the most preferred value of $d$.
Since all colluders have the same most preferred value and $d$ is
	a colluder, no colluder would benefit from any deviation.
Thus we have $d \not\in C$.

Let $\vtype'_C$ be a type vector for agents in $C$ such that all
	agents in $C$ have type $\vtype_i$.
Let $\vtype'=(\vtype'_C, \vtype_{-C})$.
Let $\vs''$ be a strategy profile same as $\vs$ except that 
	when the colluders in $C$ have type $\vtype'_C$, they pretend that
	they have type $\vtype_C$ and then use strategy $\vs'_C$.
The strategy profile $\vs''$ is still legal, because both
	$\vs$ and $\vs'$ are legal and colluders either use $\vs$
	or $\vs'$ in all cases.

We now argue that all non-faulty colluders are better off
	with failure pattern $F$ and type vector $\vtype'$.
First, consider the run in which colluders do not cheat.
The consensus decision in the run of $\vs$ with $F$ and $\vtype'$ is
	dictator $d$'s most preferred value, and since $d$ is not a colluder,
	it is the same as the consensus decision in the run 
	of $\vs$ with $F$ and $\vtype$,
	which is $v_1$.
Second, consider the run in which colluders manipulate $\vs$
	in the above described manner.
The consensus decision in the run of $\vs''$ with $F$ and $\vtype'$
	is the same as the run of $\vs'$ with $F$ and $\vtype$, according
	to the manipulation rule.
Thus this decision is $v_2$.
By our assumption above, in $\vtype'$ all colluders prefer $v_2$ over
	$v_1$, and thus as long as they do not crash, they will be better
	off with the manipulation.
This contradicts to the condition that $\vs$ is a $(c,f)$-resilient
	consensus protocol under the new definition.
\end{proof}


\addtocounter{remcnt}{1}
\paragraph{Remark \arabic{remcnt} (on the non-applicability of
	the positive results of~\cite{ADGH06,ADH08}).}
The solution concept $(k,t)$-robust
	equilibrium in~\cite{ADGH06,ADH08} is stronger than ours, yet their
	positive results do not apply to our case.
The reason is that their $t$-immune definition is stronger than
	our legal strategy profile definition in terms of fault tolerance.
In their definition, they require that no Byzantine group of size at most $t$
	could decrease the utility of any non-Byzantine agents in any
	case.
In contrast, our definition follows the standard consensus definition, and thus
	we only require that consensus is reached in spite of crash failures,
	but it is possible that crash failures change the consensus
	decision and decrease some agent's utility.
This difference in the fault tolerance requirement leads to a
	significant difference in the case with a trusted mediator.
Our $(c,f)$-resilient consensus protocol can be trivially realized
	with a trusted mediator: the mediator simply receives all proposals in
	the first round, and selects the value from the agent with the smallest
	identifier and broadcast this value as the decision value.
However, no $(k,t)$-robust equilibrium exists even with a trusted mediator.
The reason is that, if one exists, there must be a dictator as our
	Theorem~\ref{thm:dictator} also applies to $(k,t)$-robust equilibrium.
Then if the dictator is a Byzantine agent, he could always select
	some value as his most preferred value in order to decrease some
	other agent's utility, which means the protocol is not $t$-immune
	even with the trusted mediator.
Due to this reason, the positive results in~\cite{ADGH06,ADH08} cannot
	be applied to our solution concept.
This is the reason why if we remove the legal strategy profile
	requirement (the solution concept 
	is still weaker than the solution concept in~\cite{ADGH06,ADH08}),
	we cannot even have a $(1,1)$-resilient consensus protocol
	(Proposition~\ref{prop:legal}), while in~\cite{ADGH06,ADH08}, 
	a $(k,t)$-robust equilibrium could exists without additional assumptions,
	as long as $n > 3(k+t)$ and a $(k,t)$-robust equilibrium with
	a trusted mediator exists.

\addtocounter{remcnt}{1}
\paragraph{Remark \arabic{remcnt} (on $(c,f)$-resiliency vs.
	$(c,f')$-resiliency for $f'<f$).}
One subtlety of our solution concept is 
	that $(c,f)$-resiliency does not directly imply 
	$(c,f')$-resiliency for $f' < f$.
The reason is due to the risk-averse 
	legal strategy profile requirement: When the number of possible failures
	decreases, risk-averse agents need to worry less number of failure 
	patterns in guaranteeing consensus, and thus having more chances to
	manipulate the protocol.
This situation exists in general if the solution concept contains
	a risk-aversion aspect: when the possible scenario in the
	environment gets smaller ($f$ decreases in our case), risk-averse
	agents have more chance to cheat, so a previous equilibrium
	may no longer be an equilibrium.


\section{Collusion-resistant consensus protocols}
\label{sec:cons}

In this section, we first describe a deterministic $(2,f)$-resilient consensus
	protocol for any $f\le n-2$, and then adapt the protocol to a
	deterministic $(2,f)$-resilient protocol and a randomized
	$(n-1,f)$-resilient protocol for any $f\le n-1$.

\subsection{\Alg2cheater: Deterministic
	$(2,f)$-resilient consensus protocol for $f\le n-2$}\label{sec:2cheater}

The deterministic consensus
	protocol, named \Alg2cheater, consists of three components.
In the first component \AlgUpdateMG, 
	agents exchange and update status of every message
	occurred so far.
In the second component \AlgDecideD, 
	agents use the message status collected 
	in \AlgUpdateMG to determine the current dictator of the
	system, in order to decide on the dictator's most preferred value.
In the third component \AlgConsistency, 
	agents perform consistency check and execute a
	punishment strategy when detecting any inconsistency.
We index every message $m$ in a run as $(i,j,r)$, which means that $m$ is
	sent by agent $i$ to agent $j$ in round $r$.
For convenience, we
	use $\sender(m)$, $\receiver(m)$, and $\round(m)$ to denote the sender,
	receiver, and the round of $m$, respectively.

\subsubsection{Component \AlgUpdateMG}

We first describe component \AlgUpdateMG.
For each message $m$ indexed by $(i,j,r)$, each agent $p$
	records $m$ in one of the four status, \sent, \notsent, \neverknown and
\uncertain, with the following intuitive meaning:
	(a) \sent: agent $p$ knows (from all messages he has
  received) that the message $m$ was sent by $i$ successfully;
	(b) \notsent: agent $p$ knows (from all messages he has
  received) that the message $m$ was not sent successfully by $i$
	because $i$ has crashed;
	(c) \neverknown: $p$ would never know whether message $m$
	is sent successfully by $i$ or not, no matter what happens later
	in the run;
	(d) \uncertain: $p$ does not know yet whether message $m$ is sent
	successfully or not, but $p$ may know about it later in the run.

Formally, each agent $p$ maintains a variable $\MsgGraph$, the value of
	which is a function from all $(i,j,r)$ tuples to
	$\{\sent, \notsent,\neverknown,\uncertain\}$.
The value of $\MsgGraph$ of agent $p$ at the end of round $r$ is
	denoted as $\MsgGraph_{p, r}$, for $r\ge 0$ ($\MsgGraph_{p,0}$ means
	the initial value of $\MsgGraph$ at $p$).
In $\MsgGraph_{p, r}$, all messages of round $r+1$ or above have the default
	status \uncertain.
For a message $m$, we use $\MsgGraph_{p, r}(m)$ to denote the status of $m$
	recorded in $p$'s variable $\MsgGraph$ at the end of round $r$, and
	when context is clear we may simply represent it as $\MsgGraph(m)$
	or $\MsgGraph_p(m)$.
The following definition is used by the algorithm to when labeling
	a message as \neverknown.

\begin{definition} \label{def:chain}
  A sequence of messages $(m_0, m_1, \ldots, m_k)$ is called the
  \textit{message chain of $m$} in some variable $\MsgGraph$ at the end
  of round $r$ if all of the following are satisfied:
	(a) $m = m_0$;
  	(b) $\sender(m_i)=\sender(m_{i-1})$ or $\sender(m_i)=\receiver(m_{i-1})$,
	for all $1 \leq i \leq k$;
  	(c) $\round(m_i)=\round(m_{i-1})+1$, for all $1 \leq i \leq k$;
  	(d) $\MsgGraph(m_i)=\uncertain$, for all $0 \leq i < k$;
  	(e) $\round(m_k)=r$, or $\MsgGraph(m_k)\in \{\sent,
	\notsent, \neverknown\}$.
\end{definition}

\begin{algorithm}[t!]
  \caption{Component 1: \AlgUpdateMG for agent $i$}\label{alg:updateMsgGraph}
  \DontPrintSemicolon
  \LinesNumbered

  initialize $\MsgGraph_i$ such that all labels are \uncertain \;
  ${\it live}^0_i = \textrm{all agents}$ \;
    
  \BlankLine
  Phase I. Sending messages in round $k$: \;
  send $\MsgGraph_i$ to all agents in $live^{k-1}_i$ \;
  \BlankLine
  Phase II. Upon receiving messages sent to $i$ in round $k$: \;

  ${\it live}^k_i = \{j\in \Pi :
  \textrm{agent $i$ received a message from $j$ in round $k$} \}$ \;

  Let $\MsgGraph_j$ be the $\MsgGraph$ value that $i$ receives
  from agent $j$ in round $k$ \;


  \Repeat {{no message labels can be changed by the above rules}} {
    
    Let $m$ be a message from $p$ to $q$ in round $r$ such that
    $\MsgGraph_i(m) = \uncertain$, for all $p,q\in \Pi$ and $1\le r
    \le k$; update $\MsgGraph_i(m)$ using the following
    rules \label{alg:MG}
    \begin{enumerate}[itemsep=0pt, parsep=0pt, topsep=0pt]
    \item label $m$ as \sent in $\MsgGraph_i$, if and only if
      \begin{enumerate}[itemsep=0pt, parsep=0pt, topsep=0pt]
      \item $q = i$ and agent $i$ received $m$ in round $r$, or
      \item $p = i$ and $r \leq k$, or
      \item \label{item:copy1} $\MsgGraph_j(m)=\sent$ for some $j\ne i$.
      \end{enumerate}
    \item label $m$ as \notsent in $\MsgGraph_i$, if and only if
      \begin{enumerate}[itemsep=0pt, parsep=0pt, topsep=0pt]
      \item $q = i$ and agent $i$ did not receive $m$ in round $r$, or
      \item $r>1$, and
	for some $j\in \Pi$,  $\MsgGraph_i((p,j,r-1)) = \notsent$, or
	// $p$ fails to send \\ a message in round $r-1$
      \item \label{item:copy2} $\MsgGraph_j(m)=\notsent$ for some $j\ne i$.
      \end{enumerate}
    \item label $m$ as \neverknown in $\MsgGraph_i$, if and only if
      \begin{enumerate}[itemsep=0pt, parsep=0pt, topsep=0pt]
      \item (i) $r=1$ or for all $j\in \Pi$, $\MsgGraph_i((p,j,r-1))\in\{ \sent,
	\neverknown\}$, and \\ 
	(ii) every message chain of message $m$ in
	$\MsgGraph_i$ ends at some message of status \\
	\notsent or \neverknown.
	// rule similar to 1(c) and 2(c) is not needed
      \end{enumerate}
    \end{enumerate}
  }
\end{algorithm}

Algorithm~\ref{alg:updateMsgGraph} shows the pseudocode of
	\AlgUpdateMG.
In each round, agents exchange and update their $\MsgGraph$'s.
The update rule for agent $i$ is summarized in line~\ref{alg:MG}.
Rules 1 and 2 for labeling message $m$ as \sent or \notsent is
	self-explanatory. 
Rule 3 for labeling a message $m$ as \neverknown is more complicated.
Essentially, what the rule says is that if $i$ does not see that
	$\sender(m)$ has crashed in the round $\round(m)-1$ (rule 3(a)(i)), and
	all possible message chains that
	could pass the status of $m$ to $i$ end up in lost 
	messages (rule 3(a)(ii)), then
	$i$ would never know the \sent or \notsent status of $m$, no matter what
	happens later, and thus $i$ labels $m$ as \neverknown.
We say that an agent {\em learns the status} of a message $m$ if
	he updates the label of $m$ to non-\uncertain.
These message labels are important for the second algorithm component to 
	determine if an agent has obtained enough information to warrant a change
	of dictatorship.
Agent $i$ maintains the set of live agents up to round $k$ 
	in ${\it live}^k_i$, and only send messages to live agents.
This is used by the second component as one mechanism to stop cheating
	behavior.

\subsubsection{Properties of \AlgUpdateMG}

In the following we list a series of lemmas that exhibit the
properties of \AlgUpdateMG. 
For all lemmas in this section, we assume that
every agent follows the algorithm until some round $T$, and except for the
	case of crashes, no agent terminates the algorithm voluntarily at or before
round $T$, and all round numbers mentioned in these lemmas are no
larger than $T$.

\begin{definition}
  A failure pattern $F$ is consistent with a message graph
	$MG$ for $p$ at the end of round $r$,
	if in $F$ the value of variable $\MsgGraph$ of
  agent $p$ at the end of round $r$ is exactly $MG$,
	i.e. $\MsgGraph_{p, r}=MG$.
\end{definition}

\begin{lemma} \label{lem:sentnotsent}
Let $m$ be a message from $p$ to $q$ in round $r$.
If $m$ is labeled as \sent (resp. \notsent)
	by some agent $i$ in its $\MsgGraph$ variable in
	a run with failure pattern $F$,
	then $(p,q,r)\in F$ (resp. $(p,q,r)\not\in F$).
\end{lemma}
\begin{proof}
Suppose $m$ is label by agent $i$ as $\sent$ in round $k$.
By the algorithm, if $i$ labels $m$ according to rule 1(c), we can always
	trace it back to the first update of $m$ to \sent by some
        agent $j$, such that the rule used is 1(a) or 1(b) in the algorithm.
If agent $j$ applies rule 1(a), which means that $j=q$ receives
	$m$ from $p$ in round $r$, so $(p,q,r)\in F$.
If agent $j$ applies rule 1(b), then $j=p$ updates the label of $m$ in
	the end of round $k$ and $r\le k$, so $p$ is alive until the end of round
	$k$, and $(p,q,r)\in F$.

Suppose now that $m$ is label by agent $i$ as $\notsent$ in round $k$.
If $i$ labels $m$ according to rule 2(c) then we can trace it back to the
	first update of $m$ to \notsent by some agent $j$ using rule 2(a) or 2(b).
If $j$ applies rule 2(a), then $q=j$ does not receive $m$ in round $r$,
	which means $(p,q,r)\not\in F$.
If $j$ applies rule 2(b), then for some $j'\in \Pi$, $(p,j',r-1)\not\in F$, and
	thus by the constraint on $F$, we know that $(p,q,r)\not\in F$.
\end{proof}

\begin{corollary}\label{cor:sent}
Let $m$ be a message from $p$ to $q$ in round $r$.
If a message $m$ is labeled as \sent(resp. \notsent) in message
	graph $MG=\MsgGraph_{i, k}$
	in a run with failure pattern $F$, then
	no agent can have the same message labeled
  	\notsent(resp. \sent) in any round in the run with the same
        failure pattern.
\end{corollary}
\begin{proof}
If $m$ is labeled as \sent in message graph
	$MG=\MsgGraph_{i,k}$ in a run with failure pattern
	$F$, by Lemma~\ref{lem:sentnotsent} $(p,q,r)\in F$.
Then no agent can label $m$ as $\notsent$ in $F$ because again
	by Lemma~\ref{lem:sentnotsent} it would imply $(p,q,r)\not\in F$.
The case of $m$ is labeled as \notsent is argued symmetrically.
\end{proof}

\begin{lemma} \label{lem:once}
No rules in \AlgUpdateMG can be applied to any message who already
has a different non-\uncertain status.
\end{lemma}
\begin{proof}
Let $m$ be a message from $p$ to $q$ in round $r$.
Suppose agent $i$ changes the status of message $m$ in its
	$\MsgGraph$ variable from $\uncertain$
	to a label $x\in \{\sent, \notsent, \neverknown\}$ in round $k$.
First by Corollary~\ref{cor:sent}, we know that if
	$x=\sent$ (resp. \notsent), agent $i$ cannot change $m$'s label
	again to $\notsent$ (resp. \sent).
By rule 3(a) we know that $i$ cannot change the label to \neverknown if
	the label is already \sent or \notsent.
Therefore, the only case left to check is $x=\neverknown$ and $i$ changes
	the status of $m$ to \sent or \notsent.

Suppose, for a contradiction, that agent $i$ updates the status of $m$
	in a later round $k'>k$ to $\sent$ or $\notsent$.
Consider the case that $i$ updates the status to $\sent$ first.
Clearly, this update cannot be done by applying rule 1(a) and 1(b), because
	these two rules imply that agent $i$ labels $m$ to $\sent$ in round $r$,
	the round in which $m$ is sent, and before this update $m$'s label
	must be \uncertain.
Thus, $i$ makes the update using rule 1(c), which means $i$ receives a
	message from $j$ in round $k'$ that contains the \sent label for $m$.
We can follow the message chain back until we find a agent $j_1$ who
	applies rule 1(a) or 1(b) to update $m$ to \sent.
Hence we have a sequence of agents $j_1,j_2, \ldots, j_t=i$, and
	a sequence of messages $m_1,m_2, \ldots, m_{t-1}$, such that
	(a) $j_\ell$ receives message $m_{\ell-1}$
	from $j_{\ell-1}$ in round $k'+\ell - t$,
	for $\ell = 2,3,\ldots, t$; and
	(b) $j_1=p$ or $q$, and round $k'-t$ is round $r$ in which message $m$
	is sent.
By condition (a)
	above and Lemma~\ref{lem:sentnotsent}, we know that on agent $i$
	the label of message $m_\ell$ can only be \uncertain or \sent, for
	all $\ell = 1,2,\ldots, t-1$.
Now consider the particular round $k$ when $i$ updates $m$ to
	\neverknown.
Right before the update, $m$'s label is \uncertain, and the
	message sequence $m=m_0, m_1, m_2, \ldots, m_x$ with $k'+x-t = k$
	of some prefix of this sequence forms
	a message chain of $m$ by Definition~\ref{def:chain}.
However, on this message chain, the label of the last message cannot be
	\notsent or \neverknown by the above argument,
	contradicting rule 3(a) used by $i$ to update
	$m$ to \neverknown.
Therefore, we have a contradiction here.

Now consider the case that $i$ updates the status of $m$ to \notsent
	in a later round $k'>k$.
Similarly, this update cannot be done by applying rule 2(a), because this
	rule implies that before the update the status of $m$ must be
	\uncertain.
For rule 3(b), it is in conflict with one of the condition in rule 3(a),
	and thus $i$ cannot apply 3(b) either.
Therefore, we can also trace back a message chain and find an agent
	$j_1=q$ who applies 2(a) on message $m$, and all agents on this chain
	receives a message from their proceeding agent in the chain.
The argument is then the same as the case above and we can show that
	agent $i$ could not have updated $m$ to \neverknown in round $k$
	if we have such a message sequence.
This concludes our proof.
\end{proof}

By the above lemma, the status of any message $m$ in agent $i$'s
	$\MsgGraph$ variable changes at most once from \uncertain
	to not \uncertain, and when it happens,
	we say that agent $i$ \textit{learns} the status of message $m$.

\begin{lemma}\label{lem:neverknown}
  If a message $m$ is labeled \neverknown in message graph
	$MG=\MsgGraph_{p, r}$ in a run with failure pattern $F$, then
        for any $i \in \Pi$ that
        is alive at the end of round $r$ in $F$, agent $i$ cannot have the same
	message labeled \sent or \notsent in any round in the run with
	failure pattern $F$.
\end{lemma}
\begin{proof}
Suppose, for a contradiction, that there exists
        $i \in \Pi$ who is alive at the end of round $r$, such that
	agent $i$ labels $m$ as \sent or \notsent in a round $r'$.
Let $r'' = \max(r,r')+1$.
By Lemma~\ref{lem:once}, $m$'s label on $i$ at the beginning of
	round $r''$ is still \sent or \notsent.
Let $F'$ be a failure pattern such that failure behavior in $F$ and $F'$
	are the same for the first $r''-1$ rounds (i.e., $(i,j,k)\in F$
	if and only if $(i,j,k)\in F'$ for all $i,j\in\Pi$ and $k\le r''-1$), and
	$p$ is alive at the end of round $r''$ and $(i,p,r'')\in F'$.
Then in the run with $F'$, $i$ will send its $\MsgGraph$, which labels
	$m$ to \sent or \notsent, to $p$ in round $r''$.
According to rule 1(c) and 2(c) of \AlgUpdateMG, $p$ will update $m$'s label
	to \sent or \notsent after receiving the round-$r''$ message from
	$i$.
However, since $p$ already labeled $m$ to \neverknown by round $r<r''$,
	this contradicts to the result of Lemma~\ref{lem:once}.
\end{proof}


\begin{corollary}\label{cor:consistent}
In any run of \AlgUpdateMG and for any $p, q \in \Pi$, $m \in {\cal
  M}$ and round $r$, such that both $p$ and $q$ are alive at the end
of round $r$, and $\MsgGraph_{p, r}(m) \ne \uncertain$ and
$\MsgGraph_{q, r}(m) \ne \uncertain$, we have $\MsgGraph_{p, r}(m) =
\MsgGraph_{q, r}(m)$.
\end{corollary}
\begin{proof}
This is direct from Corollary~\ref{cor:sent} and Lemma~\ref{lem:neverknown}.
\end{proof}


\begin{lemma} \label{lem:transfer}
If agent $q$ receives a message from agent $p$ in round $r$, then
	all message status learned by $p$ by the end of round $r-1$ are learned
	by $q$ by the end of round $r$.
\end{lemma}
\begin{proof}
If a label $p$ learned by the end of round $r-1$ is \sent or \notsent, then
	by rule 1(c) or 2(c), it is clear that $q$ learns the label.
We now claim that for all \neverknown labels $p$ learned by
	the end of round $r-1$, $q$ learns all of them by the end of round $r$.
Let $m_0,m_1,\ldots, m_t$ be the order of messages in which
	$p$ learns their \neverknown labels.
If $p$ learns multiple \neverknown labels in the same round, the order is
	still the order in which $p$ applies rule 3(a) in this round.
We prove our claim by an induction on the message order.

In the base case, when $p$ learns the \neverknown label of $m_0$, all other
	learned labels are \sent or \notsent.
Thus when $q$ receives $p$'s message in round $r$, those \sent or \notsent
	labels will be learned by $q$.
If at this point the label of $m_0$ on $q$ is still \uncertain, we can see
	that all conditions in rule 3(a) are satisfied, because they were
	satisfied on $p$ when $p$ learned the label (a possible difference is
	that the message chain of $m_0$ on $q$ may end at a \neverknown message
	while on $p$ it must end at a \notsent message).
Therefore, $q$ will learn the \neverknown label of $m_0$ and the base
	case is correct.

For the induction step, suppose that for messages $m_0,m_1,\ldots, m_s$,
	$q$ learns their \neverknown labels by the end of round $r$.
For message $m_{s+1}$, $p$ learns its \neverknown label based on
	\sent/\notsent labels of other messages as well as the
	\neverknown labels of $m_0, m_1, \ldots, m_s$, which can all be learned
	by $q$ by the end of round $r$.
Therefore, if the label of $m_{s+1}$ in $q$ is still \uncertain,
	rule 3(a) can be applied on $q$, and $q$ will
	learn its \neverknown label at the end of round $r$.
This completes the induction step.

Therefore, we know that $q$ will learn all the labels that have learned
	by $p$ in the previous round, and the lemma holds.
\end{proof}

\begin{lemma}\label{lem:status}
If a message $m$ is labeled \sent in $\MsgGraph_{p, r}$ for
    some agent $p$ and round $r$, then for all message status that
    $\sender(m)$ has learned before round $\round(m)$, agent $p$ also
    learns them in round $r$.
\end{lemma}
\begin{proof}
We prove this lemma by an induction on the round number $r$, with base
	case $r=\round(m)$.

When $r=\round(m)$, since $p$ updates $m$'s label to \sent by the
	end of round $r$, $p$ must do so using either rule 1(a) or 1(b).
If $p$ uses 1(a), then $p=\receiver(m)$.
By Lemma~\ref{lem:transfer}, $p$ learns all the labels that $\sender(m)$
	learns before round $\round(m)$.
If $p$ uses 1(b), then $p=\sender(m)$, and the statement is trivially true.
Thus the base case is correct.

For the induction step, suppose that the statement is true for rounds
	from $\round(m)$ to $r$, and we need to prove it for round $r+1$.
If $p$ updates the label of $m$ via rule 1(a), it must have done so
	in round $\round(m)$, and the statement holds by induction hypothesis.
If $p$ updates the label via rule 1(b), the statement is trivially true.
Suppose now $p$ updates the label of $m$ via rule 1(c), in particular,
	$p$ receives a message $m'$ from $q$ in round $r+1$ in which
	$m$ is labeled as \sent.
By induction hypothesis, by the end of round $r$
	$q$ learns all the labels that $\sender(m)$
	has learned before round $\round(m)$.
Since $p$ receives message $m'$ from $q$ in round $r+1$, by
	Lemma~\ref{lem:transfer}, $p$ learns all the message status that $q$ has
	learned by the end of round $r$, which include all the message status that
	$\sender(m)$ has learned before round $\round(m)$.
Thus the induction step is also correct, and the lemma holds.
\end{proof}

\begin{lemma}\label{lem:roundk}
For any agent $i$ and round $k$, if $i$
  learns the status of all messages of round $k$, then $i$ learns the
  status of all messages before round $k$.
\end{lemma}
\begin{proof}
  Suppose, for a contradiction, that there exists some message
	before round $k$ that agent
  $i$ has not learned yet.
Among all such messages, let $m$ be the one in the earliest round,
	and let it be a message from $p$ to $q$ in round $r < k$.
Since agent $i$ has learned all
  messages of round $k$, every message chain of message $m$ must end
  at some message of or before round $k$, with status either
  \sent, \notsent or \neverknown. Consider the following cases:
  \begin{enumerate}
  \item $r > 1$ and one of the messages from agent $p$ in
    round $r-1$ is labeled \notsent.
	According to rule 2(b) of
    \AlgUpdateMG, message $m$ should be labeled as \notsent by $i$.
  \item There exists some message chain $(m, m_1, \ldots, m_k)$ of $m$
	on agent $i$
    such that the status of $m_k$ is \sent.
	By the definition of message chain of $m$, $\sender(m_k)$ is
	either $\sender(m_{k-1})$ or $\receiver(m_{k-1})$, and thus
	$\sender(m_k)$ learns the status of $m_{k-1}$ by the end of
	round $\round(m_{k-1}) = \round(m_k)-1$.
	By Lemma~\ref{lem:status}, agent $i$ learns the message status
	of $m_{k-1}$.
	This contradicts the definition of message chain, which requires that
	the label of $m_{k-1}$ is \uncertain.
  \item The rest cases. According to rule 3(a)
    of \AlgUpdateMG, message $m$ would be labeled \neverknown.
  \end{enumerate}
  Thus we reach a contradiction in every case, which proves the lemma.
\end{proof}

\begin{lemma}\label{lem:2roundlearn}
If no agents crash in round $t$
	and $t+1$ for some $t\ge 1$, then for any correct agent $i$ and a message
	$m$ of round $t$, $i$ learns the status of $m$ at
  	the end of round $t+1$.
\end{lemma}
\begin{proof}
Let message $m$ of round $t$ be indexed as $(p,q,t)$.

If $(p,q,t)\in F$, then $p$ is alive in round $t$ and $t+1$, and $p$ will
	label message $m$ as $\sent$ at the end of round $t$
	(rule 1(b)) and $i$ will
	receive this label from $p$ in round $t+1$ and label $m$ as
	\sent (rule 1(c)).

If $(p,q,t) \notin F$, then $p$ crashes in round $t-1$ or earlier.
If $m$ is still \uncertain on agent $i$ at the end of round $t+1$, consider
	a message $m'$ that is the earliest round \uncertain message on $i$
	at the end of round $t+1$.
Let $m'$ be indexed as $(p',q',r')$, with $r' \le t$.
If $r'>1$ and there exists a $j$ such that $(p',j,r'-1)$ is labeled as
	\notsent on agent $i$, then $i$ will label $m'$ as \notsent (rule 2(b)).
Thus consider either $r'=1$ or for all $j\in \Pi$,
	$(p',j,r'-1)$ is labeled as either \sent or \neverknown (cannot be
	\uncertain by the selection of $m'$), so condition (i) of rule 3(a)
	holds.

Consider any message chain of $m'$, namely $m'=m_0, m_1,\ldots, m_k$, of
	agent $i$ at the end of round $t+1$.
We argue below that either the chain ends with a label of \notsent or
	\neverknown, or $m_k$ will be removed from the chain after labeling
	$m_{k-1}$ on agent $i$.
Let $m_k$ be indexed as $(p_k,q_k,r_k)$.
\begin{itemize}
\item {\em Case 1.} $m_k$ is labeled \sent.
Since $m'=m_0$ is labeled \uncertain, we know $k\ge 1$.
By Lemma~\ref{lem:sentnotsent}, $p_k$ is alive in round $r_k-1$.
Since $p_k=\sender(m_{k-1})$ or $\receiver(m_{k-1})$,
	$p_k$ learns the status of $m_{k-1}$ by the end of round $r_k-1$.
By Lemma~\ref{lem:status}, $i$ learns the status of $m_{k-1}$ by the end
	of round $t+1$, which means that $m_k$ will be removed from the message
	chain after learning the status of $m_{k-1}$.

\item {\em Case 2.} $\round(m_k)=t+1$.
In this case, since we know that $m'$ is in round $r' \le t$, we have
	$k\ge 1$.
If $p_k$ is alive in round $t+1$, then $p_k$ learns the status of $m_{k-1}$
	at the end of round $t$, and $p_k$ will send a message to $i$ in
	round $t+1$.
By Lemma~\ref{lem:transfer}, $i$ will learn the status of $m_{k-1}$, and thus
	$m_k$ will be removed from the above message chain
	after $i$ learns the status of $m_{k-1}$.
If $p_k$ is not alive in round $t+1$, then $p_k$ crashes before round $t$.
Thus $p_k$ will not send a message to $i$ in round $t$, causing $i$ to label
	message $(p_k, i, t)$ as \notsent (rule 1(a)).
Then at the end of round $t+1$, $i$ will label $m_k$ as \notsent.
Hence in this case the chain ends with a label \notsent.

\item {\em Case 3.} $m_k$ is labeled \notsent or \neverknown.
This is what we want.
\end{itemize}
We have exhausted all cases for a message chain.
Our conclusion is that, if there are still message chains left after further
	labeling, all message chains end with labels \notsent or \neverknown.
This satisfies condition (ii) of rule 3(a).
Therefore, in this case, message $m'$ would be labeled as \neverknown.

We can repeat the above argument such that all \uncertain messages in the
	earliest round will be labeled with \sent, \notsent, or \neverknown, which
	implies that the status of message $m$ will be learned by $i$ at
	the end of round $t+1$.
\end{proof}


\begin{corollary}\label{cor:eventuallearn}
Let $t$ be the last round that some agent crashes in failure pattern $F$.
For any correct agent $i$ and any message
  $m$, $i$ learns the state of $m$ in \AlgUpdateMG in round
  no late than $r = \max\{t+2, \round(m)+1\}$ in the run with failure
  pattern $F$.
\end{corollary}

\begin{proof}
  If $\round(m) \leq t+1$, since no agent crashes in round $t+1$ and $t+2$,
	by Lemma~\ref{lem:2roundlearn}, at the end of round
  $t+2$, agent $i$ learns the status of all messages of round
  $t+1$.
Thus by Lemma~\ref{lem:roundk}, agent $i$ learns the status
  of message $m$. If $\round(m) > t+1$, because no agent crashes in
  round $\round(m)$ and $\round(m)+1$, again by
  Lemma~\ref{lem:2roundlearn}, agent $i$ learns the status of $m$ at
  the end of round $\round(m)+1$.
\end{proof}


\begin{lemma}\label{lem:neverlearn}
For any message $m$, if there exists
  some round $r \geq \round(m)$, such that $\MsgGraph_{p, r}(m) \neq
  \notsent$ for all agents $p$ that are alive at the end of round $r$, 
	then no agents can label $m$ as \notsent in any round later than
  $r$.
\end{lemma}

\begin{proof}
  Suppose by contradiction that there exists some message $m$ such
  that the statement of the lemma is not true.
If there are more than one such
  messages, we pick the one with the smallest $\round(m)$. Assume that
  some agent $p$ labels $m$ as \notsent in some round $r' > r$, and
  again we pick the smallest $r'$ such that this holds.

  First it is easy to see that agent $p$ cannot label $m$ as \notsent
  by rule 2(a). If agent $p$ learns it by rule 2(b), this means there
  exists a message $m'$ with $\round(m') = \round(m)-1$, and
  $\MsgGraph_{p, r}(m') = \notsent$.
By the assumption on the minimality of $\round(m)$, the statement of
	the lemma is true
  for $m'$, which means there exists $p' \in \Pi$, such that
  $\MsgGraph_{p', r}(m') = \notsent$, then by rule 2(b) we should have
  $\MsgGraph_{p', r}(m) = \notsent$, which is a contradiction. Finally,
  if agent $p$ learns the status of $m$ by rule 2(c), this means some
  agent learned this status in round $r'-1$, again this contradicts
  the definition of $r'$.
\end{proof}

\begin{lemma}\label{lem:alive}
If agent $i$ labels a message $m$ from $p$ to $q$ in round $r$ as
	\notsent in its variable $\MsgGraph_i$, and either $r=1$ or
	$i$ labels all messages	from $p$ in round $r-1$ as \sent or
	\neverknown, then $q$ must be alive at the end of round $r$.
\end{lemma}
\begin{proof}
By the condition that either $r=1$ or
	$i$ labels all messages	from $p$ in round $r-1$ as \sent or
	\neverknown, we know that $i$ does not apply rule 2(b) when labeling
	$m$ as \notsent.
If $i$ applies rule 2(a), then it is clear that $q=i$ is alive at the
	end of round $r$.
Suppose now $i$ applies rule 2(c), by receiving $\MsgGraph_j$ from $j$
	with $\MsgGraph_j(m)=\notsent$.
When $j$ labels $m$ to \notsent, it cannot be the case that $r>1$ and
	there exists some $j'$ such that $j$ has labeled message from
	$p$ to $j'$ in round $r-1$ as \notsent,  because if so
	$i$ would label this message as \notsent too when receiving
	$\MsgGraph_j$ from $j$, but we know that $i$ labels
	such messages either as \sent or \neverknown, contradicting to
	Lemma~\ref{lem:once}.
Hence $j$ cannot apply rule 2(b) when labeling $m$.
If $j$ applies rule 2(c) when labeling $m$, we can repeat the above argument
	again, until we track back to an agent $j_0$ who applied rule 2(a)
	when labeling $m$ as \notsent.
This means $q=j_0$ is alive at the end of round $r$.
\end{proof}



\subsubsection{Components \AlgDecideD and \AlgConsistency}

In the second component, \AlgDecideD (shown in Algorithm~\ref{alg:collude})
	maintains the current dictator of the system and decides the preferred
	value of the dictator when it is safe to do so.
Component \AlgDecideD runs in parallel with \AlgUpdateMG, which means
	that (a) when an agent $i$ sends a message in \AlgDecideD, the message would
	be piggybacked together with the message sent in \AlgUpdateMG;
	(b) \AlgDecideD
	reads some variables maintained in \AlgUpdateMG,
	in particular, $\MsgGraph_i$, and ${\it live}^k_i$; and
	(c) after receiving messages in each round, every agent first runs
	Phase II of \AlgUpdateMG and then
	runs Phase II of \AlgDecideD.

\def\standby{{\sf standby}}
\def\prene{{\sf pre-newepoch}}

\begin{algorithm}[t]
\caption{Component 2: \AlgDecideD for agent $i$ with most preferred value $v_i$}\label{alg:collude}
\DontPrintSemicolon
\LinesNumbered

${\it dictator}_i = 1$\;\label{line:firstdictator}
\BlankLine
Phase I. Sending messages in round $k$:\;

\If {$\dictator_i = i$ and agent $i$ has never decided a value before} {
  Send $\newepoch(v_i)$ to all agents in ${\it live}^k_i$ \label{line:sendne} \;
}
\BlankLine
Phase II. Upon receiving messages sent to $i$ in round $k$:\;

// First apply Phase II of \AlgUpdateMG to
	process $\MsgGraph$ messages received in this round\;
\If {agent $i$ decided a value in round $k-1$} {
  {\bf terminate} all components of \Alg2cheater \label{line:terminate} \;
}

\If {$\dictator_i = i$} {
  decide $v_i$; end Phase II \label{line:decide1} \;
}

\Repeat 
{{$\dictator_i$ does not change in the current iteration}
  \nllabel{line:dicuntil}
} {\label{line:dicrepeat}
  \If {received $\newepoch(v)$ from $\dictator_i$ at round $k' < k$ {\bf and} \\
    $\forall j\in \Pi, \MsgGraph_i(\dictator_i, j,
    k')\in\{\sent,\neverknown\}$ \nllabel{line:decidecond2} }{
    decide $v$, the most preferred value of $\dictator_i$; end Phase II
    \label{line:decide2} \;
  }
  \ElseIf {$\dictator_i\notin live^k_i$ \nllabel{line:ifchangedic} } {
    $r = \min \{r'\ |\ \exists j\in\Pi, \MsgGraph_i(\dictator_i,j,r')
    =\notsent\}$ \label{line:minr} \;
    \If {($r= 1$ {\bf or}
      $\forall j\in\Pi, \MsgGraph_i(\dictator_i,j,r-1)\ne \uncertain$) {\bf and}
      \\ $\forall j\in\Pi,\MsgGraph_i(\dictator_i,j,r)\ne \uncertain$
      \nllabel{line:newdic}} {
      $\dictator_i = \min\{j\in \Pi\ |\
      \MsgGraph_i(dictator_i,j,r)=\notsent\}$ \label{line:changedictator} \;
    }
  }
}
\end{algorithm}

In \AlgDecideD, initially all agents set agent $1$ as the default
	dictator (line~\ref{line:firstdictator}).
The current dictator $d$ sends out his most preferred value $v_d$ in
	a $\newepoch(v_d)$ message to all agents (line~\ref{line:sendne}).
If dictator $d$ is still alive at the end of the round, he simply decides 
	on $v_d$ (line~\ref{line:decide1}),
	and sends one more round of messages before terminating the algorithm
	(line~\ref{line:terminate}).

If dictator $d$ crashes before he decides, 
	other agents rely on their $\MsgGraph$'s
	to determine if they still decide on $d$'s most preferred value or
	switch to a new dictator (lines~\ref{line:dicrepeat}--\ref{line:dicuntil}).
If an agent $i$ finds out that all messages from dictator $d$ in the round when
	$d$ sends the \newepoch messages are labeled either \sent or \neverknown,
	he will decide on $v_d$ (line~\ref{line:decide2}).
This is because algorithm \AlgUpdateMG guarantees that in this case no
	alive agent can detect the crash of the dictator and it is indeed possible
	that the dictator already decides from the point of view of live agents.
If instead $i$ detects that the dictator $d$ crashes before 
	sending out all \newepoch messages, $i$ finds the first
	round $r$ in which some message from dictator $d$ is labeled 
	\notsent (line~\ref{line:minr}), and if all messages from $d$
	in round $r-1$ and $r$ have non-\uncertain labels, $i$ switches the
	dictator from $d$ to a new $d'$ who has the smallest id among all
	agents not receiving messages from $d$ in round 
	$r$ (lines~\ref{line:newdic}--\ref{line:changedictator}).
The change of dictatorship indicates that a new epoch starts, and agents
	repeat the same procedure above in determining whether to follow
	the current dictator or switch to a new one.


\begin{algorithm}[t]
\caption{Component 3: \AlgConsistency for agent $i$ with most
  preferred value $v_i$}\label{alg:consistency}
\DontPrintSemicolon
\LinesNumbered

At the end of each round $k$ // after Phase II of \AlgUpdateMG and
	\AlgDecideD: \;
Let $\mhist_i[1..k]$ be the message history of agent $i$.\;
Construct a failure pattern $F'$ such that for each message
$m$ indexed by $(p, q, r)$, $(p, q, r) \notin F'$ if and only if $\MsgGraph_i(m) = \notsent$. \label{step:fp} \label{line:constructF}\;
\If {$\exists p,q,q',r$, s.t. $(p,q,r)\in F' \wedge (p,q',r-1)\notin F'$  or
	number of crashes in $F'$ is larger than $f$ \label{line:ifvalidF}} {
//$F'$ is not a valid failure pattern \;
decide $\top$ ($\top\notin V$); {\bf terminate} all components of \Alg2cheater
\label{line:punishF} \;
}
Let $\vv' = (v'_1 = \bot, v'_2 = \bot, \ldots, v'_i
= v_i, \ldots, v'_n = \bot)$. // vector of simulated most preferred values 
	\label{line:simulateb}\;
\ForEach{$j \neq i$ such that $i$ has received a $\newepoch(v_j)$ message from $j$}{
    Let $v'_j = v_j$\;
}
Simulate \AlgUpdateMG and \AlgDecideD with failure pattern $F'$
and vector $\vv'$ of most preferred values up to round $k$. 
	\label{line:simulate}\;
Let $\mhist'_i[1..k]$ be the message history for agent $i$ in
the simulation with $F'$ and $\vv'$.\;
\If{$\mhist_i[1..k] \neq \mhist'_i[1..k]$}{
  decide $\top$; terminate all components of \Alg2cheater 
	\label{line:simulatee}\;
}
\end{algorithm}

Component \AlgConsistency (shown in Algorithm~\ref{alg:consistency}) is for
	agent $i$ to detect any inconsistency in the run due to manipulations
	by colluders.
It is run at the end of each round after the Phase II of both 
	\AlgUpdateMG and \AlgDecideD have completed.
The message history collected on agent $i$ is inconsistent if it cannot
	be generated by any valid failure pattern when following the protocol.
\AlgConsistency avoids enumerating all failure patterns by constructing
	one plausible failure pattern $F'$ from $\MsgGraph_i$ and 
	then simulating the run of \AlgUpdateMG and \AlgDecideD only in this
	failure pattern.
Once inconsistency is detected, agent $i$ decides a special value $\top$
	not in the set $V$ of possible proposals, which violates Validity
	of consensus.
This acts as a punishment strategy to deter colluders. 

\subsubsection{Proof of \Alg2cheater (\AlgUpdateMG + \AlgDecideD + 
	\AlgConsistency)}

We now proof that \Alg2cheater is a correct $(2,f)$-resilient
	consensus protocol for any $f\le n-2$.
In our analysis, we first show that for any $f\le n-1$, in order to manipulate
	\Alg2cheater, some cheater has to either pretend a crash failure or
	fake the receipt of a message he does not receive, since most other
	cheating actions are deterred by consistency checks.
Then when $f \le n-2$, pretending a crash is prevented by the second condition
	in line~\ref{line:ifvalidF} of \AlgConsistency, because it is possible
	that the total number of 
	failures exceed $f$ if a cheater pretends a crash, which
	could be detected by an honest agent.
However, guarding against faking messages is much more subtle,
	which relies on \newepoch messages and the way we change dictatorship.
To do so, we carefully design the conditions for an agent to claim
	dictatorship in \AlgDecideD, among which one important
	condition is that a new dictator 
	has to be among agents who do not
	receive the \newepoch messages from the previous dictator, which
	reduces the incentive for an agent to fake the message.
These conditions together with the properties of \AlgUpdateMG guarantee
	that the cheater has to fake the status of a \newepoch message, and
	the cheater has to crash for his colluding partner to benefit.
Finally, we consider an alterative run in which the cheater who fakes
	the message is alive but his partner dies at the same time he
	fakes the message.
We show that in this case, the lone cheater cannot get the most preferred
	value from the final dictator, the one of whom the cheater fakes the
	message, and thus the cheater has the risk of not able to decide in
	 a run.

For Lemmas~\ref{lem:dicchain} to~\ref{lem:transferchain}, we assume that
every agent follows the algorithm until some round $T$, and except for the
	case of crashes, no agent terminates the algorithm voluntarily at or before
round $T$, and all round numbers mentioned in these lemmas are no
larger than $T$.

For any agent $i$, $\dictator_i$ may change during the run.
We define \textit{dictator chain of agent $i$} to be the sequence of
	dictator values occurred in variable $\dictator_i$ of agent $i$.
Note that the first dictator in the dictator chain is $1$ according to
	line~\ref{line:firstdictator}.

\begin{lemma}\label{lem:dicchain}
Let $D_0=1, D_1, D_2, \ldots, D_t$ be a dictator chain on some agent $i$.
For every pair of consecutive dictators $D_{\ell-1}$ and $D_\ell$ in the chain,
	there exists a message $m_\ell$
	from $D_{\ell-1}$ to $D_\ell$ in some round $r_\ell$, such that
	(a) $m_\ell$ is labeled as \notsent on agent $i$,
	(b) $D_{\ell-1}$ crashes by round $r_\ell$,
	(c) $D_\ell$ is alive at the end of round $r_\ell$, 
	(d) no labels of messages from $D_{\ell-1}$ in round $r_\ell -1$ is
	labeled \notsent; and
	(e) $r_1 < r_2 < \cdots < r_t$.
\end{lemma}
\begin{proof}
When $i$ changes the dictator from $D_{\ell-1}$ to $D_{\ell}$, it does so
	because the message $m$ from $D_{\ell-1}$ to $D_{\ell}$ in some round $r$
	is labeled \notsent (line~\ref{line:changedictator}).
Let $m_\ell=m$, and $r_\ell=r$.
Condition (d) in the statement is true because $r_\ell$ is the earliest round
	in which the label of some messages from $D_{\ell-1}$ is \notsent
	(line~\ref{line:minr} and first condition of line~\ref{line:newdic}).
By Lemma~\ref{lem:sentnotsent}, $D_{\ell-1}$ crashes by round $r_\ell$.
Moreover, we know that either $r_\ell=1$ or
	messages from $D_{\ell-1}$ in round $r_\ell-1$ is labeled \sent or \neverknown
	(implied by line~\ref{line:minr} and the first condition of
	line~\ref{line:newdic}).
By Lemma~\ref{lem:alive}, we know that $D_{\ell}$ is alive at the end of
	round $k$.
Since $D_\ell$ crashes by round $r_{\ell+1}$, we know that
	$r_\ell < r_{\ell+1}$.
\end{proof}

\begin{corollary}\label{cor:uniquedictator}
Any agent can appear at most once in any dictator chain.
\end{corollary}
\begin{proof}
Immediate from Lemma~\ref{lem:dicchain} and the fact that no agent becomes
	alive again after a crash.
\end{proof}

\begin{lemma}\label{lem:prefix}
Let $D_0=1, D_1, D_2, \ldots, D_t$ and
	$D'_0=1, D'_1, D'_2,$ $ \ldots, D'_{t'}$  be two dictator chains
	on agents $i$ and $j$ respectively, at the end of some round $k$
	when they are both still alive in the run.
Then one chain is a prefix of the other chain.
\end{lemma}
\begin{proof}
We prove the result by an induction on the length of the shorter chain.
The base case of $D_0=D'_0$ is already given.
Suppose that $D_\ell = D'_{\ell}$.
When agent $i$ changes the dictator from $D_\ell$ to $D_{\ell+1}$,
	it finds a minimum round $r$ in which a message from
	$D_\ell$ has label \notsent (line~\ref{line:minr}) and no message
	from $D_\ell$ in round $r-1$ or $r$ is labeled \uncertain
	(line~\ref{line:newdic}).
Similarly, when agent $j$ changes the dictator from $D'_\ell$ to $D'_{\ell+1}$,
	it finds such a minimum round $r'$.
We claim that $r=r'$.
If not, suppose without loss of generality that $r<r'$.
By line~\ref{line:minr}, on agent $i$ message from $D_\ell$ to
	$D_{\ell+1}$ in round $r$ is labeled
	as \notsent by the end of round $k$.
By rule 2(b) of \AlgUpdateMG, message from $D_\ell$ to
	$D_{\ell+1}$ in round $r'-1$ is labeled as \notsent on agent $i$
	by the end of round $k$.
By Corollary~\ref{cor:sent} and Lemma~\ref{lem:neverknown},
	agent $j$ cannot label the message $D_\ell$ to
	$D_{\ell+1}$ in round $r'-1$ as \sent or \neverknown by the end of
	round $k$.
Since agent $j$ selects a new dictator $D'_{\ell+1}$ by the end of round $k$,
	by the first condition in line~\ref{line:newdic} of \AlgDecideD,
	$j$ cannot label the message $D_\ell$ to
	$D_{\ell+1}$ in round $r'-1$ as \uncertain either at the end of round $k$.
Thus, $j$ must have labeled this message as \notsent, but this contradicts
	the fact that $r'$ is the minimum round in which some message from
	$D_\ell$ is labeled as \notsent on agent $j$.
Therefore, we have $r=r'$.
Finally, since both $i$ and $j$ are alive by the end of round $k$,
	by Corollary~\ref{cor:consistent}, $i$ and $j$ have the same
	non-\uncertain labels for all messages from $D_\ell$, and thus
	they must have selected the same dictator $D_{\ell+1}$.
\end{proof}

Note that the requirement that both $i$ and $j$ are alive in round $k$ cannot
	simply be removed, and here is a simple counter-example if it is removed.
Consider a system of four agents.
In round $1$ agent $1$ crashes and fails to send messages to agents $2$ and $4$.
In round $2$, agent $2$ crashes and fails to send a message to $4$.
In round $3$, agent $3$ crashes and fails to send a message to $4$.
All other messages are successfully sent.
We can check that at the end of round $2$ agent $3$ will change the dictator
	from $1$ to $2$, but at the end of round $3$ agent $4$ will change
	his dictator from $1$ to $4$, because he will label the message from
	$1$ to $2$ in round $1$ as \neverknown at the end of round $3$.
Therefore, the dictator chain $1,4$ on agent $4$ at the end of round $3$
	and the dictator chain $1,2$ on agent $3$ at the end of round $2$
	are not prefix of each other.

\begin{corollary} \label{cor:singleNE}
No two agents can send $\newepoch(\cdot)$ messages in the same round.
\end{corollary}
\begin{proof}
Suppose, for a contradiction, that both agents $i$ and $j$ send
	$\newepoch(\cdot)$ in the same round $k$.
By the algorithm, it is clear that $k$ cannot be $1$, and
	by the end of round $k-1$, $i$ is the last dictator in the dictator
	chain on $i$, and $j$ is the last dictator in the dictator chain
	on $j$.
By Lemma~\ref{lem:prefix}, either $i$ appears in the chain of $j$ or
	the reverse is true.
Suppose $i$ appears in the chain of $j$.
By Lemma~\ref{lem:dicchain}, $i$ must have crashed by a round $r$ and
	$r < k-1$, which contradicts to our assumption that $i$ sends
	out $\newepoch(\cdot)$ in round $k$.
\end{proof}

\begin{lemma} \label{lem:transferchain}
If agent $j$ receives a message from agent $i$ in round $k+1$, then
	the dictator chain of agent $i$ at the end of round $k$ is a prefix
	of the dictator chain of agent $j$ at the end of round $k+1$.
\end{lemma}
\begin{proof}
By Lemma~\ref{lem:prefix}, at the end of round $k$, either the dictator chain of
	$i$ is a prefix of dictator chain of $j$, or the reverse is true.
If the dictator chain of $i$ is a prefix of dictator chain of $j$ at
	the end of round $k$, then of course it is also a prefix of the
	dictator chain of $j$ at the end of round $k+1$.
Suppose now that the dictator chain of $j$ at the end of round $k$ is a prefix
	of the dictator chain of $i$.
By Lemma~\ref{lem:transfer}, at the end of round $k+1$ $j$ learns the status
	of all messages that $i$ learn at the end of round $k$.
Then the condition that causes $i$ to change dictators by round $k$ would also
	cause $j$ to change dictators at the end of round $k+1$.
Therefore, $j$ will change the dictators exactly as $i$'s chain,
	and make its dictator chain at least
	as long as $i$'s at the end of round $k$.
Thus the lemma holds.
\end{proof}

\begin{lemma}\label{lem:terminate}
  Suppose that all agents are honest. When an agent terminates the
  algorithm at line~\ref{line:terminate} of \AlgDecideD in some round
  $k$, all agents that are still alive in round $k$
	must all have decided by the end
  of round $k$.
\end{lemma}

\begin{proof}
  Let agent $i$ be the first agent to terminate the algorithm from
  line~\ref{line:terminate} of \AlgDecideD, in some round $k$.
Let $D$ be the last
  dictator on agent $i$ who sends out \newepoch message at some round
  $k'$. Suppose agent $i$ decides on $D$'s most preferred value in
  round $k''$. According to algorithm, agent $i$ terminates the
  algorithm at the end of round $k''+1$, i.e. $k=k''+1$.
Note that no agent terminates
  the algorithm before round $k$, which means all lemmas from
  Lemma~\ref{lem:sentnotsent} to Lemma~\ref{lem:transferchain} hold
	by the end of round $k$.
Then for any agent $p$ that are still alive in round $k$, he must
	have received
  agent $i$'s \MsgGraph variable in round $k$. By
  Lemma~\ref{lem:transferchain}, agent $p$ must have agent $i$'s
  dictator chain as a prefix at the end of round $k$. Because
  $\MsgGraph_i(D, j, k') \in \{\sent, \neverknown\}$ for all $j \in
  \Pi$, by Lemma~\ref{lem:transfer}, these status are also learned by
  agent $p$ by the end of round $k$. According to the
  algorithm, agent $p$ will decide on agent $D$'s most preferred value
  by the end of round $k$.
\end{proof}

With Lemma~\ref{lem:terminate}, we know that no agents voluntarily terminate
	the algorithm before all other agents decide.
Therefore, the voluntary termination
	of the algorithm by any agent
	does not affect the decisions of the agents, and
	we can apply all previous lemmas until the termination of the algorithm.

\begin{lemma} \label{lem:decidebot}
  Suppose that all agents are honest. No agent can reach
  line~\ref{line:punishF} in \AlgConsistency and decide $\top$.
\end{lemma}

\begin{proof}
  First it is easy to see that when all agents are honest, the $F'$
  constructed in \AlgConsistency must be a valid failure pattern,
  i.e., if any agent $i$ fails to send a message to some agent in
  $F'$, $i$ will not send any message to any agent in the next round.
  Thus the first condition in line~\ref{line:punishF} cannot be true
  at any time.

  According to the algorithm, once an agent decides through
  line~\ref{line:decide1} or line~\ref{line:decide2} of \AlgDecideD,
  he will terminate the algorithm before reaching
  line~\ref{line:punishF} in \AlgConsistency in the next round. Thus
  it suffices to prove that no agent can have $|{\it live}_i| < n - f$
  at any round when or before he decides.

  Suppose that some agent $p$ has $|{\it live}^t_p| < n - f$ at some
  round $t$. And among all such cases, we pick the one with the
  smallest round number $t$. This means that there are more than $f$
  agents who do not send messages to agent $p$ in round $t$. Since at
  most $f$ agents can crash in a run, there must be some agent $i$ who
  terminates the algorithm at line~\ref{line:terminate} in some round
  earlier than round $t$. Then by Lemma~\ref{lem:terminate}, all other
  alive agents have also decided before round $t$. Thus the lemma
  holds.
\end{proof}

\begin{lemma}\label{lem:consistent}
  Suppose that all agents are honest. No agent can reach
  line~\ref{line:simulatee} in \AlgConsistency and decide $\top$.
\end{lemma}

\begin{proof}
  Given any failure pattern $F$ and most preferred value vector $\vv$,
  for any agent $i$ and any round $k$. Let $F'$ and $\vv'$ be the
  failure pattern and most preferred value vector that agent $i$
  constructed in \AlgConsistency. Let $\MsgGraph$ and $\MsgGraph'$ be
  the variables in the run with $(F, \vv)$ and $(F', \vv')$,
  respectively.

  First, for any $j \in \Pi$ and $r \leq k$, because message $(j, i,
  r)$ is labeled either \sent or \notsent in $\MsgGraph_{i, k}$,
  according to the construction rules at Step~\ref{step:fp}, we know
  $(j, i, r) \in F$ if and only if $(j, i, r) \in F'$. In the
  following we will prove that for any $(j, i, r) \in F$ and any
  message $m$, we have $\MsgGraph_{j, r}(m) = \MsgGraph'_{j, r}(m)$.
  \begin{itemize}
  \item If $\MsgGraph_{j, r}(m) = \sent$, which implies that
    $\MsgGraph_{i, k}(m) = \sent$. Hence $m \in F$ and $m \in F'$. If
    in the run with failure pattern $F$, this status is labeled by
    agent $j$ according to rule 1(a) or 1(b) of \AlgUpdateMG, then $j$
    can also label it as \sent using the same rule in $\MsgGraph_{j,
      r}$ with $F'$. If $j$ labels it by rule 1(c) with failure
    pattern $F$, we can follow the message chain back until we find
    some agent $j'$ who applies rule 1(a) or 1(b) to update $m$ to
    \sent. Then with failure pattern $F'$, $j'$ will apply the same
    rule to label $m$ to \sent, and every agent in the message chain
    will label $m$ as \sent, as they did with $F$, and eventually $j$
    will label $m$ as \sent in $\MsgGraph'_{j, r}$. On the other hand,
    if $\MsgGraph'_{j, r}(m) = \sent$, using a similar argument, we
    can show that $\MsgGraph_{j, r}(m) = \sent$. This means that
    $\MsgGraph_{j, r}(m) = \sent$ if and only if $\MsgGraph'_{j, r}(m)
    = \sent$.
  \item Using the similar argument as in previous case, we can show
    that $\MsgGraph_{j, r}(m) = \notsent$ if and only if
    $\MsgGraph'_{j, r}(m) = \notsent$.
  \item Let $M$ be the set of all messages with status \neverknown in
    $\MsgGraph_{j, r}$ with failure pattern $F$, sorted by the time of
    them being labeled. For any $m \in M$, if all messages before it
    in $M$ are all labeled \neverknown in $\MsgGraph'_{j, r}$ with
    failure pattern $F'$. Then rule 1(c) can also be applied to this
    message $m_i$ so that it can be labeled as \neverknown in
    $\MsgGraph'_{j, r}$ with failure pattern $F'$ too. And by
    induction, we have $\MsgGraph'_{j, r}(m) = \neverknown$ for all $m
    \in M$. Again, by a similar argument, one can show that if
    $\MsgGraph'_{j, r}(m) = \neverknown$, we have $\MsgGraph_{j, r}(m)
    = \neverknown$.
  \end{itemize}
  Finally, having the above three results, we can directly have that
  for any message $m$, $\MsgGraph_{j, r}(m) = \uncertain$ if and only
  if $\MsgGraph'_{j, r}(m) = \uncertain$. Hence, we conclude that all
  the \MsgGraph variables that agent $i$ receives in each round are
  the same with failure pattern $F$ and $F'$.

  Notice that whether an agent sends out a $\newepoch$ message in some
  round only depends on his \MsgGraph variable in that round. Also if
  agent $i$ receives $\newepoch(v_j)$ from some $j$ with most
  preferred value vector $\vv$, we must have $v'_j = v$ in $\vv'$ too,
  which means if $j$ sends out a \newepoch message with most preferred
  value vector $\vv'$, it should also be $\newepoch(v_j)$. Thus agent
  $i$ will receive the same set of $\newepoch$ messages with $(F,
  \vv)$ and $(F', \vv')$. Hence, we can conclude that $\mhist[1..k] =
  \mhist'[1..k]$. This completes the proof of this lemma.
\end{proof}

\begin{lemma}[Termination] \label{lem:termination}
Suppose that all agents are honest.
Every correct agent eventually decides.
\end{lemma}
\begin{proof}
Suppose, for a contradiction, that some correct agent $i$ does not decide.
By Corollary~\ref{cor:uniquedictator}, an agent appears at most once in the
	dictator chain of $i$, so eventually the dictator on $i$ does not change
	any more.
Suppose the last dictator on $i$ is $D$.
Since dictator only has a finite number of changes, it is clear that
	agent $i$ will not loop forever in the repeat-until
	loop (lines~\ref{line:dicrepeat}--\ref{line:dicuntil}) in any round.
Since $i$ does not decide, we know that $D\ne i$, otherwise, $i$ would
	decide in line~\ref{line:decide1}.

Suppose first that $D\notin {\it live}^k_i$ for some round
	$k$.
Note that the reason that $D\notin {\it live}^k_i$
	could be either that $D$ crashes, or that $D$ has terminated its
	\AlgUpdateMG and \AlgDecideD tasks, but we do not need to distinguish
	these two cases here.
Then $\MsgGraph_{i,k}(D,i,k) = \notsent$.
By Corollary~\ref{cor:eventuallearn}, $i$ eventually
	learns the status of all messages by round $k$.
According to lines~\ref{line:minr} and~\ref{line:newdic},
	$i$ would change the dictator in line~\ref{line:changedictator},
	contradicting to the assumption that $D$ is the last dictator on $i$.

Now suppose that $D\in {\it live}^k_i$ for all rounds $k$.
This means that $D$ is a correct agent and $D$ does not terminate his
	\AlgUpdateMG and \AlgDecideD tasks.
Therefore, $D$ will receive a message from $i$ after $i$ already fixes its
	dictator chain.
By Lemma~\ref{lem:transferchain}, after $D$ receives this message,
	$i$'s dictator chain will become a prefix of $D$'s dictator chain,
	which means that $D$ is in $D$'s dictator chain.
Then after $D$ sets itself as the dictator, $D$ must successfully send
	$\newepoch$ messages to all live agents in a round $r$ and then decide
	in line~\ref{line:decide1}.
Agent $i$ must have received this $\newepoch$ message in round $r$ from $D$.
By Corollary~\ref{cor:eventuallearn}, eventually $i$ learns the status
	of all messages from $D$ in round $r$.
Since $D$ is a correct agent, for all $j\in \Pi$, $(D, j, r)\in F$, where $F$
	is the failure pattern of the run.
Thus, by Lemma~\ref{lem:sentnotsent}, $i$ cannot label any message from $D$
	in round $r$ as $\notsent$.
According to the condition of line~\ref{line:decidecond2}, $i$ will decide
	$D$'s most preferred value ($i$ knows this value because $i$ receives
	$D$'s $\newepoch$ message containing the value).
This contradicts our assumption that $i$ does not decide.

We have discussed all cases, all of which lead to a contradiction.
Therefore, the lemma is correct.
\end{proof}


\begin{lemma}[Uniform Agreement] \label{lem:agreement}
Suppose that all agents are honest.
No two agents (correct or not) decide differently.
\end{lemma}

\begin{proof}
First, since all agents are honest, by Lemma~\ref{lem:decidebot} and
        Lemma~\ref{lem:consistent} no agent decides $\top$ in
        \AlgDecideD or \AlgConsistency.
Thus, no matter whether an agent decides in line~\ref{line:decide1} or
	line~\ref{line:decide2} of \AlgDecideD, the agent always decide the most preferred value
	of the current dictator.

  Let agent $i$ be the agent that decides in the earliest round among
  all agents. Suppose agent $i$ decides at the end of round $r$. And
  let $D$ be the last dictator on $i$ who sends out \newepoch message
  in round $k \le r$. By algorithm we know that $\MsgGraph_{i, r}(D, j, k) \in
  \{\sent, \neverknown\}$ for all agents $j$.

  For any agent $p \neq i$ who is still alive at the end of round $r$
  (otherwise he cannot decide according to our assumption), consider
  the message $(D, p, k)$.
First we know $\MsgGraph_{p, r}(D, p, k)
  \neq \neverknown$, since $k \le r$ so either $p$ receives the message
	from $D$ in round $k$ and label it as $\sent$, or $p$ does not
	receive the message and label it as $\notsent$.
Second, because $\MsgGraph_{i, r}(D, p, k) \in
  \{\sent, \neverknown\}$, by Corollary~\ref{cor:sent} and
  Lemma~\ref{lem:neverknown}, agent $p$ cannot have message $(D, p,
  k)$ labeled $\notsent$. Thus agent $p$ must have received agent
  $D$'s \newepoch message at the end of round $k$.
Notice that agent
  $D$ must have labeled herself as the last dictator when sending out the
  \newepoch message.
By Lemma~\ref{lem:transferchain}, agent $p$ must
  also have agent $D$ in the dictator chain after he received $D$'s
  message at the end of round $k$. Again since $\MsgGraph_{i, r}(D, j,
  k) \in \{\sent, \neverknown\}$ for all agents $j$, by
  Corollary~\ref{cor:sent} and Lemma~\ref{lem:neverknown}, none of
  these messages can be labeled \notsent in agent $p$'s \MsgGraph
  variable in any round. Thus agent $p$ cannot change his dictator
  from agent $D$ to any other agents, which means if he decides, the
  decision value must be agent $D$'s most preferred value too.
\end{proof}

\begin{lemma}\label{lem:consensus}
\Alg2cheater solves consensus problem if all agents are honest.
\end{lemma}
\proof
Validity is trivial.
Termination and Uniform Agreement are proven by Lemmas~\ref{lem:termination}
	and~\ref{lem:agreement}, respectively.
\endproof

We say that a cheater {\em pretends a crash} if he stops send messages
	to some honest agents in a round $r$ and then stop sending all messages
	to all honest agents in
	all rounds after $r$.
We say that a cheater {\em fakes a message} if he does not receive a message
	from an agent $p$ in a round $r-1$ but he labels this message as
	\sent in his message to all honest agents in round $r$.
The following lemma applies to any size of colluding groups, and shows that
	pretending a crash and faking a message are something a cheater has to
	do if he wants to manipulate the system.
\begin{lemma}\label{lem:cheatcases}
If any group of cheaters
	can strategically manipulate the protocol \Alg2cheater in a run, then
	some cheater must either pretend a crash or fake a message in the run.
\end{lemma}
\begin{proof}
By our model, a cheating agent $i$ may change his algorithm $A_i$, which
	given a round number $r$, a message history $\mhist[1..r]$, and
	his private type $\theta_i$, outputs the messages $\smsgs$ to be sent in
	the next round $r+1$ and a possible decision value $d$ (perhaps $\bot$).
Thus $i$ would either change the output $d$ or the messages $\smsgs$ to be
	sent.

We prove the lemma by the following case analysis on the possible cheating
	behavior of the cheating agents.
\begin{itemize}
\item
{\em Case 1.}
No cheater changes the message output in any round, and only some
	cheating agent changes the decision output $d$.
Suppose a cheater $i$ is the first who
	changes his decision output $d$ at the end of round $r$.

\begin{itemize}

\item
{\em Case 1.1.}
By the end of round $r$, some honest agent already decides.
In this case, $i$ cannot change the decision value, since otherwise it
	would violate Uniform Agreement of consensus specification.

\item
{\em Case 1.2.}
Some honest agent $j$ is in ${\it live}^r_i$, i.e., $i$ still receives
	a message from $j$ in round $r$.
Then there exists a failure pattern extension consistent with what $i$
	observes so far in which $j$ is a non-faulty agent.
By the Termination property, $j$ must decide a value $d_j$.
By agreement, all cheaters including $i$ may only decide the same value
	$d_j$, so we have $d=d_j$, that is $i$ has to decide on the value
	$d_j$ too.
Since no cheater changes their message output of their algorithms,
	agent $i$ would decide $d_j$ if he follows the protocol, therefore
	$i$ cannot benefit by changing his decision output in this case.

\item
{\em Case 1.3.}
No honest agent decides and all honest agents fail to send a message to
	$i$ in round $r$.
In this case, at the end of round $r$, if $\dictator_i$ is still an honest
	agent, according to
	lines~\ref{line:ifchangedic}--\ref{line:changedictator}, $i$ would
	change the dictator to be one of the cheaters.
Thus, $i$ would eventually decide on cheaters' most preferred value
	(their preferred values are all the same by our model assumption) if
	$i$ follows the protocol, so
	$i$ cannot benefit by deviating from the protocol.
\end{itemize}
By the above argument, we know that Case 1 cannot happen if cheaters manipulate
	the system and benefit.

\item
{\em Case 2.}
At least one cheater changes some round message.
If cheaters only change messages to other cheaters, then they do not affect
	the behavior of honest agents, and following the same argument as in
	Case 1 they will not benefit with such cheating actions.
Thus, suppose cheater $i$ changes his round $r$ message to another
	honest agent $j$.
\begin{itemize}
\item
{\em Case 2.1.} Agent $i$ is supposed to send a message to $j$ in round $r$
	but it drops this message.
If $i$ does not receive $j$'s round-$r$ message, then either $j$ crashes
	or $j$ already terminates his algorithm.
In the former case, $i$ drops a message to a crashed agent and thus it
	has no effect to the protocol outcome.
In the latter case, $j$ already decides and $i$ cannot change the decision
	anyway.
Thus $i$ does not benefit if $i$ does not receive $j$'s round-$r$ message.
Now suppose that $i$ receives $j$'s round-$r$ message.
In this case, $i$ has to stop sending messages to all honest agents in
	round $r+1$.
Otherwise, if $i$ still sends a message to some honest agent $p$ in
	round $r+1$, it is possible that $j$ would also send a message to
	$p$ in round $r+1$ in which $j$ would label the
	round-$r$ message from $i$ to $j$ as $\notsent$ (by rule
        2(a)).
Then $p$ would receive a message from $i$ in round $r+1$, and at the
same time $p$ knowns from agent $j$ that $i$ did not send $j$ a message
in round $r$. Since there is no failure pattern in which both of these two
things happen at the same time, which means agent $p$ will detect
inconsistency in \AlgConsistency and decide $\top$, violating Validity
of consensus. By the same argument, $i$ has to stop sending all
messages to all honest agents in any round after round $r$.
That is, $i$ must pretend a crash, which matches the first case
	covered in the statement of the lemma.

\item
{\em Case 2.2.}
Agent $i$ is supposed to send a message to $j$ in round $r$, but $i$
	changes the message into a wrong format.
This will cause $j$ to detect inconsistency in \AlgConsistency
	and decide $\top$, violating Validity of consensus.
Thus this case cannot occur.

\item
{\em Case 2.3.}
Agent $i$ is supposed to send a $\newepoch$ message in round $r$ to $j$ but
	it does not send it (it still sends the $\MsgGraph$ message), or
	it is not supposed to send a $\newepoch$ message but it sends such a
	message.
Since whether to send a \newepoch message or not can be derived from the
$\MsgGraph_i$ variable of agent $i$ at the beginning of round $r$,
this cheating behavior cannot be act alone, otherwise
	$j$ will detect an inconsistency in \AlgConsistency and decide $\top$, violating
	Validity of consensus.
Thus the cheaters must also cheat in some other way.

\item
{\em Case 2.4.}
Agent $i$ is supposed to send $j$ $\newepoch(v)$ in round $r$ but
	instead he sends $j$ $\newepoch(v')$.
If agent $i$ consistently changes all his $\newepoch(v)$ messages to
	$\newepoch(v')$, this is equivalent of $i$ changing his private type,
	if there is no other cheating actions combined.
However, when fixing any failure pattern, it is clear that our protocol
	is a dictatorship protocol, meaning that it always decides on some
	agent's most preferred value.
Thus cheaters cannot benefit by changing their private type in a dictatorship
	protocol.
If agent $i$ changes his $\newepoch(v)$ messages inconsistently, namely
	sending different values to different honest agents, then if it is not
	combined with other cheating actions (such as pretend a crash or
	fake a message label), it is possible that $i$ is correct,
	and two honest agents would decide two different values, violating
	Uniform Agreement of consensus.

\item
{\em Case 2.5.}
Agent $i$ changes some message labels in the $\MsgGraph$ he sends to
	$j$ in round $r$.
Without loss of generality, we could assume that this is the earliest
	label-cheating action among
	all such label cheating actions.
Among all message labels that $i$ cheated, let $m$ be the message of
	the latest round.

\begin{itemize}
\item
{\em Case 2.5.1.}
Message $m$ is of round $r-2$ or earlier.
Then $m$'s label sent by $i$
	in round $r-1$ must be \uncertain, because otherwise $i$ changes
	$m$'s non-\uncertain label from round $r-1$ to round $r$, and $j$
	would detect the inconsistency and decide $\top$.
The fact that
	$i$ labels $m$ as uncertain at the end of round $r-2$ implies that
	$i$ is neither the sender nor the receiver of $m$, because otherwise
	$i$ has to label
	$m$ either $\sent$ or $\notsent$ by the end of round
	$r-2$ according to the algorithm.

Suppose sub-case A is
	that $i$ is supposed to label $m$ as some non-\uncertain label
	at the end of round $r-1$, but $i$ changes the label.
That means $i$ got enough information in round $r-1$ allowing $i$ to
	apply rule 1(c), 2(b), 2(c), or 3(a) of \AlgUpdateMG on $m$.
However, in this case, it is possible that agents who provide these information
	also provide the same information to $j$ (recall that no one cheat
	message labels in round $r-1$), and also $j$ receives correct labels
	from $i$ in round $r-1$ mapping to the \MsgGraph state of $i$ in the
	end of round $r-2$, so $j$ could have label $m$ to the same non-\uncertain
	label at the end of round $r-1$.
Thus, $i$ cannot change $m$'s label to different non-\uncertain label,
	since otherwise $j$ would detect inconsistency.
If $i$ cheats $m$'s label to be \uncertain, then it must also pretend that
	it does not receive enough information in round $r-1$, which means
	$i$ has to cheat on labels of some messages sent to $i$ in round
	$r-1$, but this contradicts our assumption that $m$ is the latest round
	message that $i$ cheats on.

Suppose now the sub-case B is that $i$ is supposed to label $m$
	as \uncertain at the end of round $r-1$, but $i$ cheats the label to
	some non-\uncertain label.
Since $j$ could have received the same information as $i$ received in round
	$r-1$, $i$ has to pretend that he receives more information from another
	agent that $i$ actually does not receive a message from, to avoid $j$
	detecting an inconsistency.
However, this means that $i$ also needs to cheat the status of a message
	in round $r-1$, contradicting to our assumption that $m$ is the latest round
	message that $i$ cheats on.

The above shows that Case 2.5.1 is not possible.

\item
{\em Case 2.5.2.}
Message $m$ is of round $r-1$.
If $i$ is the sender of the message, $i$ has to
	follow the algorithm and label $m$ as \sent in his
	round $r$ message, because any agent can detect consistency if he
	labels $m$ to something else.
If $i$ is not the sender nor the receiver of the message, $i$ has to follow
	the algorithm and
	label $m$ as \uncertain, again because any agent can detect consistency
	otherwise.
Thus let $i$ be the receiver of $m$.

Consider first that $i$ receives $m$, but cheats $m$'s label as $\notsent$.
If $\sender(m)$ is an honest agent, it is possible that $\sender(m)$
	is able to send a round $r$ message to $j$, and then $j$ will detect
	an inconsistency.
If $\sender(m)$ is also a cheater, it has to drop the message to $j$ in order
	to avoid inconsistency.
This goes back to Case 2.1, as we conclude that $\sender(m)$ has to pretend
	a crash.

Finally, consider that $i$ does not receive $m$, but cheats $m$'s
	label as $\sent$ in his message to $j$ in round $r$.
In this case, $i$ has to consistently send the \sent label of $m$
	to all live and
	honest agents in the round $r$, because otherwise, two honest agents
	may exchange message in round $r+1$ and detects inconsistency in the
	labeling of $m$ (some labels $m$ as \sent while others labels
	$m$ as \notsent).
This is exactly the faking message case in the statement of the lemma.
\end{itemize}
\end{itemize}
\end{itemize}
We have exhausted all cases, and show that in all runs some cheater has
	to either pretend a crash or fake a message to in order to
	benefit.
\end{proof}

Note that Lemma~\ref{lem:cheatcases} does not preclude cheaters to use
	other cheating actions such as deciding earlier, but it dictates that
	cheaters have to combine other cheating methods with pretending a crash
	or faking a message to be successful.
With this lemma, to show that the protocol is collusion-resistant, it is
	enough to show that these two cheating actions cannot occur in any run.

\begin{lemma}\label{lem:pcrash}
In \Alg2cheater, no group of agents of size at most two
	can strategically manipulate protocol by some cheater pretending
	a crash, when $f < n-1$.
\end{lemma}

\proof
By Lemma~\ref{lem:cheatcases}, we only need to show that cheaters cannot
	pretend a crash or fake a message.
We first show that cheaters cannot pretend a crash.
Suppose, for a contradiction, that a cheater $i$ pretends a crash in
	round $r$ by not sending a message to an honest agent $j$.
Suppose that there are $\ell$ agents crashed before round $r$.
In round $r$, we can crash another $f-\ell$ agents including the other cheater
	so that none of them sends messages to agent $j$ (meaning finding
	another failure pattern that satisfies these conditions).
For agent $j$, it will detect at the end of round $r$ that
	$|{\it live}^r_i| < n-f$ and decide $\top$
        (line~\ref{line:punishF} in \AlgConsistency).
This means consensus is violated and the strategy profile is not legal.
Thus cheaters cannot pretend crashes when $f < n-1$.
\endproof

\begin{lemma}\label{lem:fakemsg}
In \Alg2cheater, no group of agents of size at most two
	can strategically manipulate protocol by some cheater faking
	a message.
\end{lemma}
\proof
Suppose, for a contradiction, that cheater $i$ fakes a message in round $r$
	by labeling a message from $p$ in round $r-1$
	as \sent while $i$ does not receive this message, and sending this label
	to all honest agents.
Without loss of generality, we assume no faking message behavior by
	any cheater has occurred in round $r-1$ or earlier.
We prove this case through the following series of claims.

Claim 1. In the run where $i$ fakes the round-$(r-1)$ 
	message from $p$ to $i$ in round $r$ and some cheater benefits from
	this cheating behavior, $p$ must be in the dictator 
	chain of any agent who decides. 

Proof of Claim 1. Suppose that $p$ is not on the dictator chain.
If $i$ does not fake the message, the dictator chain would remain the same,
	and thus no cheater can benefit from this cheating behavior, 
	a contradiction. 

Claim 2. In some run in which $i$ conducts the above cheating action, 
	$p$ is alive at the end of round $r-2$.

Proof of Claim 2. Since $i$ pretends that $p$ sends a message
	to $i$ in round $r-1$, and this cheating behavior does not cause
	any agent to detect inconsistency, there must be a run in which 
	$p$ indeed is alive at the end of round $r-2$ and sends a message
	to $p$ in round $r$.

Henceforth, we consider a run $R$ in which 
	$p$ is alive at the end of round $r-2$, $i$ fakes the round-$(r-1)$ 
	message from $p$ to $i$ in round $r$, and some cheater benefits from
	this cheating behavior.

Claim 3. 
In the $\MsgGraph$ sent out by $i$ in round $r$, 
	all round-$(r-2)$ messages addressed to $p$ are labeled
	\sent or \notsent, and the labels match the failure pattern of
	run $R$.

Proof of Claim 3. If $i$ would successfully receive the message from 
	$p$ in round $r-1$, this message would contain \sent or \notsent labels
	of all messages addressed to $p$ in round $r-2$, and $i$ cannot
	fake these labels in his round-$r$ message because $p$ may have successfully
	sent these labels to other honest agents, who might be able to detect
	inconsistency if $i$ does so.
Note that $p$ itself could be a cheater, but by our assumption $p$ does
	not cheat in round $r-1$.

Claim 3 means that at the end of round $r-1$ agent $i$ knows the status
	of all messages addressed to $p$ in round $r-2$.

Claim 4. For any agent $q$, if $p$ does not receive $q$'s message in
	round $r-2$, then $q$ must have failed to send out some message
	in round $r-3$, and $i$ labels this message as \notsent at the
	end of round $r-1$.

Proof of Claim 4. By Claim 3, agent $i$ would label the message from
	$q$ to $p$ in round $r-1$ as \notsent, which correctly
	matches the failure pattern.
Thus, according to Algorithm \AlgUpdateMG, if $i$ were an honest agent,
	$i$ could only apply rule 2(b) for this message, which means
	some message from $q$ in round $r-3$ or earlier is labeled \notsent
	by $i$, and indeed $q$ fails to send this message.

Claim 5. For any agent $q$, if $p$ does not receive $q$'s message in
	round $r-2$, then at the end of round $r-2$, $p$ must have labeled
	some message from $q$ in round $r-3$ and
	all messages from $q$ in round $r-2$ as \notsent.

Proof of Claim 5. By Claim 4, $q$ must have failed sending a message $m$
	in round $r-3$ and $i$ labels $m$ as \notsent by round
	$r-1$.
If agent $i$ learns this label at the end of round $r-3$, then $i$
	would pass this label to $p$ in round $r-2$.
If agent $i$ learns this label in round $r-2$ or later, then this label
	must be passed to $i$ through a chain of messages. 
Let message from $x$ to $y$ in round $r-2$ be a message on this chain.
Then agent $x$ must have also passed the \notsent label of $m$ to $p$
	in round $r-2$.
This is because, if $x$ fails to send a message to $p$ in round $r-2$, 
	by Claim 4 $x$ should have crashed by round $r-3$ and thus cannot
	send a message to $y$ in round $r-2$.
Therefore, $p$ would learn the \notsent status of $m$ at the end of 
	round $r-2$.
Since $m$ is a message of round $r-3$, $p$ would label all messages
	from $q$ in round $r-2$ as \notsent.

Claim 6. by the end of round $r-2$, agent $p$ has learned the status 
	of all messages of round $r-3$ or earlier.

Proof of Claim 6. Consider an arbitrary message $m$ from $x$ to $y$ in round
	$r-3$.
If $p$ receives a message from $x$ or $y$ in round $r-2$, then $p$ would learn
	the status of $m$.
Suppose that $p$ does not receive messages from $x$ and $y$ in round
	$r-2$.
Then by Claim 5 $p$ would label all messages from $x$ and $y$ in round
	$r-2$ as \notsent.
According to rule 3(a) of \AlgUpdateMG, in this case $p$ would label
	$m$ as \neverknown.
By Lemma~\ref{lem:roundk}, $p$ would learn the labels of all messages by
	round $r-3$.

Claim 7. In run $R$, at the end of round $r-2$, $p$ 
	sets itself as the dictator.

Proof of Claim 7. By Claim 1, $p$ must be on the dictator chain.
If $p$ is the first on the dictator chain, the claim is trivially true.
If not, let $d$ be the dictator before $p$ in the dictator chain.
According Lemma~\ref{lem:dicchain}, there exist a message $m'$ of round $r'$
	from $d$ to $p$, 
	such that $p$ is alive at the end of round $r'$ and $d$ fails to
	send $m'$ to $p$.
Since $p$ crashes in round $r-1$, we know that $r'\le r-2$.
If $r'=r-2$, then by Claim 5, $p$ would have labeled some message from $d$
	in round $r-3$ as \notsent, which
	contradicts condition (d) of Lemma~\ref{lem:dicchain}.
If $r'\le r-3$, by Claim 6, $p$ learns the status of all messages by
	round $r'$.
According to Algorithm \AlgDecideD, $p$ should have changed the dictator
	from $d$ to $p$ at the end of round $r-2$.

Claim 8. Agent $p$ would send the $\newepoch(v_p)$
	messages to all agents in round $r-1$.

Proof of Claim 8. If agent $p$ knows that he is the dictator by
	round $r-3$, then $p$ would decide at the end of round $r-2$, and
	no one can change the decision any more.
Thus, $p$ must know his dictatorship at the end of round $r-2$.
According to Algorithm \AlgDecideD, $p$ would send the $\newepoch(v_p)$
	messages to all agents in round $r-1$.

Claim 9. Agent $i$ crashes in run $R$.

Proof of Claim 9. Suppose that $i$ does not crash. 
By Claim 8, $i$ does not receive the $\newepoch(v_p)$ message from $p$
	in round $r-1$.
If in run $R$ agent $i$ does not become a dictator after $p$, then there is no
	effect for $i$ to fake the message from $p$ to $i$ in round $r-1$.
If $i$ indeed becomes the next dictator, since $i$ does not crash, 
	$i$ would decide on his most preferred value, and thus $i$ will not
	benefit from cheating.
Thus, $i$ must eventually crash in run $R$.

Claim 10. Agent $p$ must be an honest agent.

Proof of Claim 10. If $p$ is also a cheater, then together with Claim 9
	we know that both cheaters $p$ and
	$i$ crash in run $R$.
Since we only have two cheaters, and by our definitions cheaters' utility
	is fixed to zero in failure patterns in which they crash, 
	in run $R$ no cheater will benefit, contradicting the definition of
	$R$.
Note that this is the place where we use the condition of $c=2$.
If $c=3$, $p$ actually could be a cheater, and manipulation behavior
	exists (see an example in Section~\ref{sec:counter2}).

We are now ready to reach the final contradiction.
Consider a run $R'$, such that
	(a) $R'$ is the same as run $R$ up to round $r-2$; 
	(b) in round $r-1$ agent $p$ successfully sends his $\MsgGraph$ and
	$\newepoch(v_p)$ messages to all agents but $i$; 
	(c) in round $r$ the other cheater (if exists) crashes without
	sending out any messages; and
	(d) $i$ does not crash in $R'$.
At the end of round $r-1$, the message history of $i$ is the same
	in two runs, so $i$ could still cheat in $R'$ as in $R$.
Let $i$ do so in $R'$.

Even though we crash the other cheater in $R'$, 
	run $R'$ is still a run with at most $f$ crash failures, because
	by Claim 9 $i$ crashes in $R$ but $i$ is correct in $R'$.
In $R'$, agent $i$ does not receive $v_p$ directly from $p$, and 
	he will not receive $v_p$ from any other honest agents according to
	the protocol, and he will not receive $v_p$ from the other cheater since
	the other cheater crashes at the beginning of round $r$.
By Claim 10 agent $p$ is an honest agent.
Therefore, in $R'$ $i$ will not know the value of $v_p$.

In $R'$ no message from $p$ in round $r-1$ can be labeled \notsent
	by any agent other than $i$, and $i$ himself cheats this label to be \sent.
According to the algorithm, in this case the final decision in run $R'$
	must be $v_p$, the most preferred value of $p$.
However, $i$ does not receive $v_p$ from any agent, and thus he cannot
	decide, a contradiction.
\endproof

\def\thmnewepoch{
Protocol \Alg2cheater is a $(2,f)$-resilient consensus protocol
	for any $f \le n-2$.
}
\begin{theorem}\label{thm:newepoch}
\thmnewepoch
\end{theorem}

\begin{proof}
The result is directly obtained from
	Lemmas~\ref{lem:consensus}, \ref{lem:cheatcases},
	\ref{lem:pcrash} and~\ref{lem:fakemsg}.
\end{proof}

\subsubsection{Counter-example for \Alg2cheater when $f=n-1$}
\label{sec:counter1}

\begin{figure}[t]
 \centering
   \includegraphics[width=0.5\textwidth]{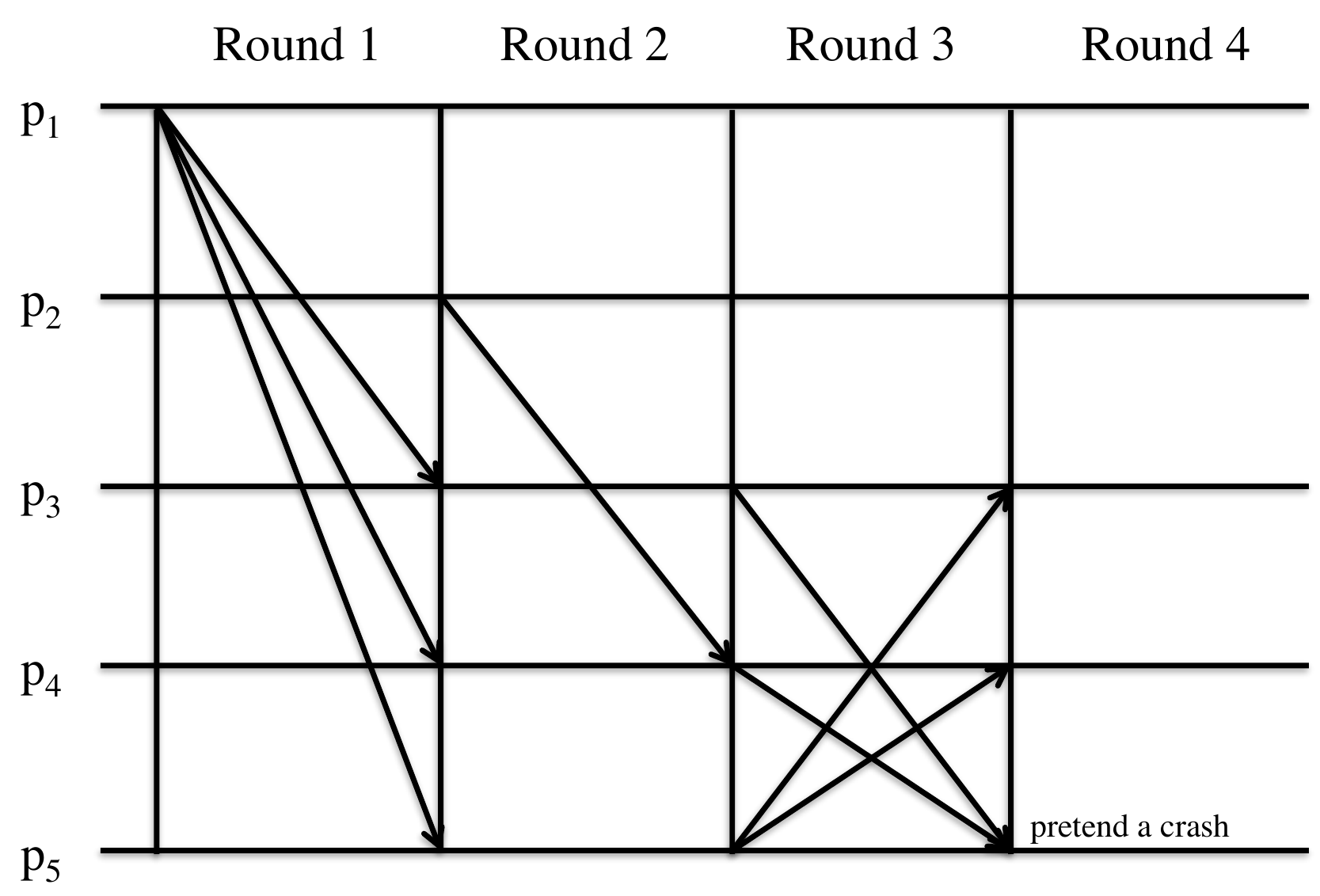}
 \caption{Counter-example of \Alg2cheater when $f=n-1$.}\label{fig:counter1}
\end{figure}
Figure~\ref{fig:counter1} describes an counter-example \Alg2cheater
when $f=n-1$. To make the example clear, it only shows the messages 
sent by $p_1$ in round $1$, messages sent by $p_2$ in round $2$ and
messages sent or received by $p_5$ in round $3$. Here, $p_4$ and $p_5$
are cheaters. In round $1$, the first dictator fails to send \newepoch
to $p_2$, but $p_2$ only tells $p_4$ about it in round $2$ and then
crashes. 
Let $m$ be the message from $p_1$ to $p_2$ in round $1$.
Then, in round $3$, $p_5$ sets $m$ to be \notsent and knows that $p_3$
sets it to be \uncertain in the end of round $2$. Note that $p_5$
does not know the status of $m$ in $p_3$'s \MsgGraph in the end of
round $3$ since $p_5$ does not know whether $p_4$ sends message to
$p_3$ in round $3$. In round $4$, $p_5$ pretends a crash and does not
send any message to $p_3$. There are several cases for $p_5$ to consider:

\begin{enumerate}
  \item In round $4$, $p_3$ sends a message to $p_5$ and tells $p_5$ that $p_4$ does not send message to $p_3$ in round $3$. Then $p_5$ decides $p_1$'s proposal value. In this case $p_5$ can benefit if $p_1$'s proposal value is better than $p_3$'s proposal.
  \item In round $4$, $p_3$ sends a message to $p_5$ and tells $p_5$ that $p_4$ sends a message to $p_3$ in round $3$, and $p_4$ does not send message to $p_5$ in round $4$. Then, $p_3$ must send a \newepoch message with his proposal
	in round $4$ since $p_3$ knows all status of messages sent by $p_2$ 
	in round $2$. 
Then, $p_5$ learns the proposal of $p_3$ and 
	decides $p_3$'s proposal value.
Note that a key point here is that even though $p_5$ pretends a crash
	at the beginning of round $4$, he is still able to receive the round-$4$
	message from $p_3$, which contains the critical information of
	$p_3$'s proposal value.
  \item In round $4$, both $p_3$ and $p_4$ send messages to $p_5$. By the same argument, $p_3$ must send \newepoch message. $p_5$ waits until round $5$. If $p_4$ sends message to $p_5$ and tells it $p_3$ fails to send \newepoch message to $p_4$ in round $4$, then $p_5$ decides $p_4$'s proposal value, otherwise 
	$p_5$ decides $p_3$'s proposal value. Note that
	it requires $p_4$ to be a cheater, otherwise $p_4$ will not send messages 
	to $p_5$ when it knows that $p_5$ has crashed.
\end{enumerate}

\subsection{\AlgWF: Deterministic $(2,f)$-resilient consensus protocol for
	$f\le n-1$ }
\label{sec:WF}

\Alg2cheater is not $(2,n-1)$ resilient as shown by the counter-example
	in Section~\ref{sec:counter1}, because 
	simply counting the number of crash failures cannot deter the
	manipulation of pretending crash failures any more.
We adapt \Alg2cheater to a new protocol \AlgWF to deal with this issue.
The only difference in the new protocol 
	is the following: When the dictator finds that
	he can send the $\newepoch$ message, he splits
	his most preferred value into two parts and sends them in two 
	consecutive	rounds with two $\newepoch$ messages separately, 
	and he decides the value by the end of the second round. Other
agents can recover the dictator's most preferred value if and only
if he knows the content of both $\newepoch$ messages. 
Note that here we assume that the proposal values need at least two bits
	to represent, which is consistent with our assumption that
	$|V|\ge 3$.

The above change, together with the requirement that agents stop sending 
	messages
	to crashed agents, successfully guards against pretending crash
	manipulations.
Intuitively, the risk when a cheater pretends a crash is that he may
	miss the \newepoch message from a new dictator and thus cannot decide
	on the most preferred value of the new dictator.
In \Alg2cheater it is possible that the cheater receives this \newepoch
	message in the same round as he pretends a crash, making him safe.
However, in \AlgWF, the most preferred values are split into two parts,
	and our analysis shows that the cheater would miss the second part if
	he pretends a crash, effectively defeating this cheating behavior.


\begin{lemma}\label{lem:nocrash}
Suppose that $1\le f \le n-1$.
  In Algorithm \AlgWF, if no cheater fakes any messages,
	then no group
  of agents of any size
	can strategically manipulate the protocol by
	some cheater pretending a crash.
\end{lemma}

\begin{proof}
First, when $f< n-1$, Lemma~\ref{lem:pcrash} can be applied to
	\AlgWF with the same proof, and thus we only consider the case
	of $f=n-1$.

Suppose, for a contradiction, that there exist a failure pattern $F$ in
  which some agent $p$ can strategically manipulate the protocol by
  pretending a crash in round $r$. Assume that agent $p$ does not
  send message to some honest agent $h$ in round $r$ and then stop
  sending messages to all honest agents in all rounds after $r$.

  Let $d$ be the last agent from which agent $h$ has received a \newepoch
  message before round $r$ (including himself), and let $k$ be the
  round in which $d$ sends its first round \newepoch message. 
Thus $k \le r-1$. 
If there
  is no such agent, let $k = 0$. Now we focus on the message status of
  all \newepoch messages send from agent $d$ in round $k$ and $k+1$
  (i.e., all \newepoch messages sent from $d$), and consider the
  following scenarios (the scenarios listed below may overlap, but
  they cover all possible cases):
  \begin{enumerate}
  \item $d = h$. If $h$ has finished sending his $\newepoch$ messages
	by round $r-1$, then $h$ will decide in round $r$ 
	and no one can change the decision.
	If $h$ is still sending his $\newepoch$ messages in round $r$,
	then only the status of these $\newepoch$ messages may cause the
	change of dictator.
Thus, whether agent $p$ sent $h$ a message or
    not in round $r$ cannot affect the final decision. 
In this case, we let $p$ send $h$ a
    message in round $r$ and reconsider the scenario.

  \item \label{case:no} $k = 0$. This means agent $h$ has never
    received a \newepoch message from any agents (including himself).
    Now consider the following failure pattern $F'$: in the first
    $r-1$ rounds, $F'$ is the same as $F$. Then all agents except $p$
    and $h$ crash at the beginning of round $r$, without sending out
    any messages. And agent $p$ and $h$ are correct agents. Because
    $F'$ is consistent with $\MsgGraph_{p, r-1}$ for agent $p$ at the
    end of round $r-1$, which means $p$ should pretend to crash in
    $F'$ as he does in $F$. Since agent $h$ will never receive any
    \newepoch messages from other agents in $F'$, she will eventually
    send out her own \newepoch messages in some round $r'$ and $r'+1$
    (with $r' \geq r$) and then decide on her own most preferred
    value. Notice that agent $p$ does not send $h$ any message in
    round $r$, thus $h$ will not send $p$ any message in any round
    later than $r$. This means $p$ can never receive the \newepoch
    messages from agent $h$ in round $r'+1$, nor can he get this
    information from other agents (because they are all crashed at the
    beginning of round $r$). Therefore, $p$ does not know what value
    does agent $h$ decide on in this case and thus are not able to
    cheat.

Note that this is the case where we require that the most preferred value
	of $h$ be splitted into two rounds $r'$ and $r'+1$, since we can only
	guarantee that $p$ does not receive the second part of the value in
	round $r'+1$.
If $h$ were to send 
	its entire proposal in round $r'$, then in the case of $r'=r$,
	$h$ would send his proposal to $p$ in round $r$ since $h$ has not
	detected that $p$ has crashed, and $p$'s cheating would be successful.
This is exactly the case shown in the counter-example in 
	Section~\ref{sec:counter1}.

  \item $k = r-1$, or there exists agent $q \in \Pi$ which is alive at
    the end of round $r-1$, and some \newepoch message $m$ sent from
    $d$ is labeled as \notsent\ in $\MsgGraph_{q,r-1}$. In the
    following we show that in this case, there always exists a
    \newepoch message $m'$ from agent $d$ and a failure pattern $F'$
    consistent with $\MsgGraph_{p, r-1}$ for $p$ at the end of round
    $r-1$, such that $\MsgGraph_{h, r}(m') = \notsent$ in $F'$. If this is
    true, we crash all other agents except $p$ and $h$ at the beginning
	of round $r+1$ and apply a similar argument as in Case~\ref{case:no}
	to show that $h$ will be the final dictator but $p$ does not know
	the most preferred value of $h$.
Thus, $p$ cannot cheat in this case.

 We consider the following subcases:
    \begin{enumerate}
    \item $k = r-1$. In this case, let $F'$ be a failure pattern
	(consistent with $\MsgGraph_{p, r-1}$ for $p$ at the end of
	round $r-1$) such that $d$ crashes in round $r$ and 
	does not send a message to $h$.
Then in $F'$ $h$ will label the message from $d$ in round $r$ as
	\notsent.

    \item $q \neq p$. Let $F'$ be the failure pattern in which agent
      $q$ send the status of $m$ to $h$ in round $r$. Thus we have
      $\MsgGraph_{h, r}(m) = \notsent$ in the run with failure pattern
      $F'$.
    \item $q = p$ and agent $p$ learns the status of $m$ before round
      $r-1$. Then $p$ must have sent this information to agent $h$ in
      round $r-1$, and agent $h$ should also labels $m$ as \notsent in
      $\MsgGraph_{h, r}$ in failure pattern $F$.
	Note that we assume that $p$ does not fake any messages in the run.

    \item $q = p$ and agent $p$ learns the status of $m$ from some
      agent $q' \ne d$ in round $r-1$. Then there exist a failure
      pattern $F'$ consistent with $\MsgGraph_{p, r}$, in which $q'$
      also sent this information to agent $h$ in round $r-1$. Hence we
      have $\MsgGraph_{h, r}(m) = \notsent$ in $F'$.

    \item $q = p$ and agent $p$ is the (supposed) receiver of this
      \newepoch message $m$ and does not receive it from agent $d$ in
      round $r-1$. In this case, $p$ does not know whether agent $d$
      has sent the \newepoch message to $h$ successfully, which means
      there is a consistent failure pattern $F'$, in which agent $h$
      does not receive the \newepoch message from $d$ neither.
    \end{enumerate}

  \item \label{case:other} $k < r-1$, and for any agent $q$ that is
    alive at the end of round $r-1$ and any \newepoch message $m$ sent
    from $d$, $\MsgGraph_{q, r-1}(m) \neq \notsent$. In this case, by
    Lemma~\ref{lem:neverlearn}, we know that no agent can label any of
    agent $d$'s \newepoch message as \notsent in any later rounds.
    Thus according to algorithm, the final decision value will be
    agent $d$'s most preferred value regardless of whether agent $p$
    crashes. Hence agent $p$ cannot benefit by pretending a crash.
  \end{enumerate}

  Above are all the possible cases. And we showed that in neither of
  them can agent $p$ cheat. This finishes the proof.
\end{proof}

\def\thmWF{The protocol \AlgWF is a $(2,f)$ resilient consensus protocol
	for any $f\le n-1$.
}
\begin{theorem} \label{thm:WF}
\thmWF
\end{theorem}

\begin{proof}
First, Lemma~\ref{lem:cheatcases} can be applied to \AlgWF with the same
	proof, which means in \AlgWF some cheater must either pretend a 
	crash or fake a message in order to benefit.
Second, Lemma~\ref{lem:fakemsg} can also be applied to \AlgWF with the same
	proof, which means no cheater can fake any messages.
Finally, Lemma~\ref{lem:nocrash} states that no cheater can pretend crashes
	when cheaters do not fake messages.
Together, we show that cheaters have no valid cheating actions, and thus
	the statement of theorem holds.
\end{proof}

\subsubsection{Counter-example for \Alg2cheater when $c=3$}
\label{sec:counter2}

\begin{figure}[t]
 \centering
   \includegraphics[width=0.5\textwidth]{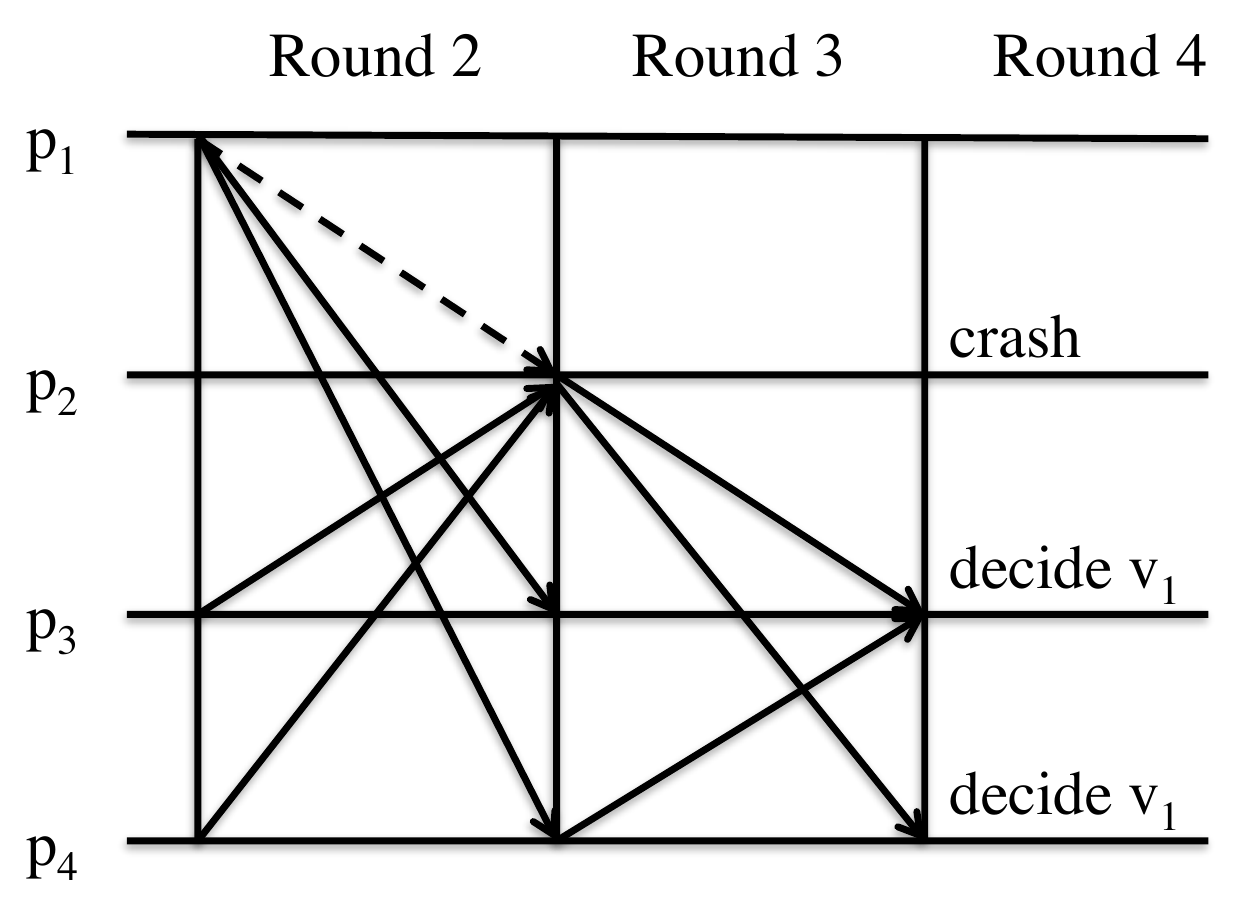}
 \caption{Counter-example for \AlgWF when $c=3$.}\label{fig:counter2}
\end{figure}
Figure~\ref{fig:counter2} describes a possible cheating example for
protocol \AlgWF when there are $3$ colluders. Let the colluders be
$p_1, p_2$ and $p_4$. Suppose in round 1, no agent crashes and agent
$1$ successfully sends his first round \newepoch messages to all other
agents. In round $2$, $p_1$ crashes and fails to send his second round
\newepoch message only to $p_2$. No other agents crash in round 2.
Then in round 3 and later rounds, $p_2$ can fake the message $p_1$
sent to him in round 2. It can be verified that no agent will detect
any inconsistency and the final decision value of the system will
always be $p_1$'s most preferred value if $p_2$ cheats in this way. In
the following, we describe a scenario where the colluders actually
benefit from this cheating behavior.

Consider the failure pattern in which $p_2$ crashes at the beginning
of round 4, without sending out any messages. Agent $p_3$ and $p_4$
are correct agents. According to algorithm, if all agents are honest,
the final decision value should be $p_3$'s most preferred value. But
as we discussed above, if $p_2$ fakes the message that $p_1$ sent to
him in round 2, the final decision will be $p_1$'s most preferred
value, and thus $p_4$ can benefit from this cheating behavior, since
	$p_4$'s most preferred value is the same as $p_1$'s.

We need three colluders here because of the following two reasons.
First, when $p_2$ fakes the receipt of message sent by $p_1$ in round $2$,
	he runs the risk that this causes the final decision to be 
	$p_1$'s most preferred value, and thus if $p_1$ were an honest agent,
	$p_2$ would not know his most preferred value and cannot decide
	(if his colluder also crashes before telling him the value).
Hence both $p_2$ and $p_1$ have to be cheaters.
Second, for cheaters to benefit, $p_2$ has to crash, since otherwise
	$p_2$ would be the next dictator and there is no need to cheat.
Since both $p_1$ and $p_2$ have crashed in the run, there has to be the
	third cheater left to take the benefit.

This example shows a scenario that a colluder has to crash in order 
	for other cheaters to benefit.
This shows the subtlety involved when our model allows crashes on colluders.

\subsection{\AlgRand: Randomized $(n-1,f)$-resilient consensus protocol
	for any $f\le n-1$} \label{sec:rand}

\AlgWF is not resilient to three colluded cheaters as shown by a counter-example
	in Section~\ref{sec:counter2}, because the colluders
	could successfully 
	manipulate the protocol through faking message receipts.
In this section, we show that if we allow agents to use randomness in
their algorithms, we can boost the protocol to resist $n-1$ colluders.

A randomized protocol is one in which every agent has access
	to random bits as part of his local state.
A randomized consensus protocol is $(c,f)$-resilient if it solves consensus
	in a system with at most $f$ crash failures regardless of random bits
	used by agents, and any strategic manipulation by at most $c$ colluders
	would lead to violation of consensus with high probability.

The randomized algorithm \AlgRand is a further adaptation of \AlgWF.
\AlgRand maintains the same structure of \AlgWF, except that messages
	are associated with random numbers as follows.
Every message is associated with a random number created by the message
	sender.
The $\MsgGraph_i$ of agent $i$ keeps track of the random number of every message
	that $i$ knows.
When $i$ sends a message $m$ to $j$ in a round $r$, 
	he first generates a copy $MG$ of $\MsgGraph_i$ in which all random numbers
	associated with messages he sends are removed.
Then he generates a unique random number $\rho_m$ associated with $m$, and
	sends $MG$ and $\rho_m$ together to $j$.
When $i$ receives a message $(\MsgGraph_p,\rho_p)$ 
	from an agent $p$ in round $r$, $i$ records $\rho_p$ in its \MsgGraph.
The consistency check in \AlgConsistency is revised as follows.
When agent $i$ simulates any message $(\MsgGraph_p,\rho_p)$ sent by an 
	agent $p$ (line~\ref{line:simulate} of \AlgConsistency), 
	if $p=i$ then $i$ uses the original random number he generated in
	the real run as $\rho_p$;
	if $i$ has received the random number from $p$ in the real run, he
	uses the received number as $\rho_p$; otherwise, he leaves $\rho_p$
	as $\bot$.
Finally, we add an additional round at the beginning for agents to exchange
	$\MsgGraph$'s and random numbers but no \newepoch messages.
Using randomness effectively stops faking message receipts with
	high probability because
	cheaters do not know the random bits used in advance.

\def\thmrand{
Protocol \AlgRand is a randomized $(n-1,f)$-resilient consensus protocol
	for any $f\le n-1$.
}
\begin{theorem} \label{thm:rand}
\thmrand
\end{theorem}

\begin{proof}
If all agents are honest, it is obvious that no agent can find any
inconsistency about the random number attached with each message.
Since the remaining part of the new algorithm is the same as the old
algorithm. Hence Lemma~\ref{lem:consensus} still holds here, i.e., the
new algorithm solves consensus problem if all agents are honest.

Using the same argument, it can be verified that
Lemma~\ref{lem:cheatcases} and Lemma~\ref{lem:nocrash} also hold for
the new algorithm. Hence in order to prove that no group of agents
can strategically manipulate the system, it is adequate to show that
no agent can fake any messages in the system.

Suppose that a cheater $p$ does not receive a message from an agent
$q$ in some round $r$. If $r = 1$, which means this is the extra round
that we added to the algorithm, then whether faking this message or
not will not effect the outcome of the algorithm. If $r > 1$, pick an
honest agent $h$ that is still alive at the end of round $r$.
Let $m$ be
the message that $h$ sends to $q$ in round $r-1$ and let $\rho$ be the
random number $h$ associates with message $m$. Note that if $p$ wants to fake
the message that $q$ sends to $p$ in round $r$, in $p$'s round $r+1$ message
	to $h$, $p$ has to include $\rho$ as the random number associated with 
	message $m$,  otherwise $h$ will
notice this inconsistency and decide on $\top$. 
However, by the end of
round $r$, agent $p$ does not know value 
	$\rho$, which means with high probability he
	cannot fake this message. This completes the proof.
\end{proof}

\subsubsection{An example that $(c,f)$-resiliency does not imply
  $(c,f')$-resiliency for $f' < f$}\label{sec:counter3}

We use protocol \AlgRand to show an example that $(c,f)$-resiliency 
	does not imply  $(c,f')$-resiliency for $f' < f$.
To be clear we use \AlgRand($f$) to denote the actual protocol with
	parameter $f$.
Consider a system of $5$ agents, $c=3$, $f=4$, and $f'=2$.
Agents $2$, $4$ and $5$ are	colluders.
Theorem~\ref{thm:rand} shows that \AlgRand($4$) is $(3,4)$-resilient
	in this system.
Note that with $f=4$, the second
	condition in line~\ref{line:ifvalidF} of \AlgConsistency
	is always true and thus useless.
We show that \AlgRand($4$) is not $(3,2)$-resilient by providing
	the following manipulation scenario.

Consider a system with at most $f'=2$ crash failures, and $f'$ is common
	knowledge to all agents.
When agent $2$ fails to receive a message from agent $1$ in round $1$, 
	agent $2$ immediately pretend a crash in round $2$ without sending out
	any messages.
We argue that this manipulation is safe to the colluders.
First, pretending a crash will not be detected by consistency check.
Second, since $f' < c=3$, and agent $1$ already crashes, 
	agent $2$ is sure that one of the remaining
	colluders $4$ or $5$ must be alive, in which case agent $2$ can always
	get the decision value from the alive colluder.
Therefore, consensus can always be achieved.

We now describe a scenario where colluders benefit.
Notice that agent $1$ is the first dictator in the run.
Suppose that agent $1$ successfully sends his round-$1$ messages to
	all other agents except agent $2$ before agent $1$ crashes in round $1$.
If agent $2$ pretends a crash at the beginning of round $2$, all other agents
	would eventually label the message from agent $1$ to agent $2$ in round
	$1$ as \neverknown.
According to our algorithm, all other messages from agent $1$ in 
	round $1$ will be labeled either as
	\sent or \neverknown, and all agents eventually decide on agent $1$'s
	most preferred value $v_1$.
However, if agent $2$ follows the protocol, and there is no
	crash failure in round $2$, at the beginning of round $3$ agent $2$
	would become the new dictator and start sending his \newepoch messages.
In this case, if agent $2$ crashes in round $3$ without sending a message
	to agent $3$, and this is the last crash failure in the run, 
	agent $3$ would become the final dictator and the decision value
	of the run would be agent $3$'s most preferred value $v_3$.
If colluders $4$ and $5$ prefers $v_1$ over $v_3$, then 
	they would benefit from agent $2$ pretending the crash in round $2$.

Therefore, we have that \AlgRand($4$) is $(3,4)$-resilient
	but not $(3,2)$-resilient.
The key is that, when there are more possible failures, agent $2$ who
	pretends a crash has the risk that all his partners crashes and
	he cannot get the final decision value in all cases, but when
	the number of possible failures decreases, he does not have this
	risk any more.

\subsection{Protocol complexity and summary of collusion resistance techniques}
\label{sec:tech}

We now discuss the complexity of the protocols, and then summarize
	a number of techniques we used in defending against strategic manipulations
	used in our protocols.

\paragraph{Protocol complexity.}
Let $f' \le f$ be the actual number of crashes in a run.
For \Alg2cheater, each crash failure delays the decision for at most $2$
	rounds by causing a change of dictatorship.
Thus, it takes at most $2f'+1$ rounds for all agents to decide and
	$2f'+2$ rounds for all agents to terminate the algorithm.
For \AlgWF, each crash delays the decision for at most $3$ rounds
	since it sends two rounds of \newepoch messages.
Its round complexity is therefore $3f'+2$ for decision and $3f'+3$ for
	termination.
\AlgRand only needs one more round than \AlgWF.
For message complexity, at most $O(n^2f')$ messages are exchanged, with
	the size of each message at most $O(n^2f')$ due to the size of \MsgGraph.
It is also easy to check that local computation on each agent is polynomial
	in $n$ and $f'$.

\paragraph{Summary on techniques for resisting strategic manipulations.}

Our protocols employ a number of techniques defending against 
	strategic manipulations, which may find applications in other
	situations.

Consistency check combined with a punishment strategy 
	(deciding $\top$ in our case) builds the first line of defense. 
It effectively restricts the possible manipulations of a cheater.
A particular form of consistency check is to count the number of
	observed failures and check if it exceeds the maximum number of
	possible failures $f$.
When $f < n-1$, this check stops agents from pretending a crash failure,
	one of the important forms of strategic manipulations.

However, consistency check is far from enough.
For example, when $f=n-1$, pretending crash failures cannot be detected.
In this case, we use the techniques of 
	not sending any messages to a crashed
	agent and  splitting critical information (the most preferred proposal
	in our case), to achieve the effect that cheaters may risk not
	receiving the critical information should they pretend a crash.

Besides pretending crash manipulations, we know from our analysis that
	another important form of manipulation is for a cheater to fake 
	a message that he does not receive. 
When sufficient random bits are available to all agents, we can let all
	agents to attach a unique sequence of random bits to each message, and
	by checking the random bits received as part of consistency checks, 
	faking messages can be prevented.
However, when random bits are not available, defending against faking
	messages is much more difficult, especially when there are colluding agents.
Our protocol \Alg2cheater combines several techniques to guard against
	faking message manipulations.
One technique is the maintenance of consistent message status through
	the message graph exchange component \AlgUpdateMG.
The second technique is carefully designed conditions for claiming
	new dictatorship, in particular, the new dictator is selected among
	agents who detect the failures of the old dictator, and this reduces
	the incentive of faking message receipts from the old dictator.


While our protocols are designed specifically for solving the problem
	of synchronous consensus,
	we believe that the above mentioned techniques could be potentially used
	in other distributed protocol design for defending against
	strategic manipulations.

\section{Impossibility of resisting colluders with private  
	communications}
\label{sec:imp}

In this section, we consider a modified synchronous round model 
	in which colluders can communicate with one another
	through private communication channels after one synchronous round
	ends but before the next round starts.
These private channels provide new opportunities for colluders to
	manipulate the protocol.
For example, if cheater $i$ receives a message from $p$ but
	cheater $j$ does not receive a message from $p$ in the same round $r$, 
	$i$ and $j$ through their private communication would know that
	$p$ crashes in this round, and thus in round $r+1$ 
	it is safe for $i$ to pretend not
	receiving a message from $p$ in round $r$, a case not feasible
	without private communication.
Indeed, our theorem below shows that with private communication
	no $(2,f)$-resilient consensus protocol exists for any $1\le f\le n-1$,
	even with randomness.

\def\thmimp{
In a synchronous system with private communication channels among
	colluders, there is no randomized $(2,f)$-resilient consensus protocol
  with $n \geq 3$ agents, for any $1\le f \le n-1$.
}
\begin{theorem} \label{thm:imp}
\thmimp
\end{theorem}

\begin{proof}
  Suppose, for a contradiction, that such a protocol exists. 
By Theorem~\ref{thm:dictator} for any failure pattern $F$ 
	the protocol has a dictator under $F$.

  Suppose without loss of generality
	that agent $1$ is the dictator in the failure-free
  run. 
Now if change the failure pattern by 
	deleting the messages sent by agent 1 one by one, round
  by round starting from the final round before agent $1$ decides, 
	there must exist a failure pattern $F$ where the dictator is still agent
  1, but if we delete (any) one more message sent from agent 1, the
  dictator will change. 
Then in failure pattern $F$, let round $r$ be the last round
	in which agent 1 has
  sent at least one message. We consider two cases here:
  \begin{itemize}
  \item[(1)] Agent 1 sends only one message in round $r$. Suppose agent
    $1$ sends a message only to agent 2 in round $r$.

    First we show that if this message is removed, the new dictator
    can only be agent 2. Assume otherwise that the dictator becomes another
	agent, say agent 3. 
Then we consider the case that agents 1 and 2 are in the
    colluding group. 
If in round $r$ agent 2 does not receive a
    message from agent 1, he could cheat by pretending that he has
    received this message. In this case, the algorithm will choose
    agent 1 as the dictator instead of agent 3, which will benefit agent 2.

    Now we consider another case that agent 2 and some other agent,
    say agent 3, are in the colluding group. Notice that agent $2$
	receives a message from agent $1$ but 
	agent 3
    does not receive a message from agent 1 in round $r$, thus through
	their private communication at the end of round $r$, both
    agent 2 and agent 3 will know that agent 1 crashes in this round.
    Hence agent 2 could pretend that he did not receive the message
    from agent 1 in round $r$. 
In this case, the dictator will become agent 2 instead of
    agent 1, which benefits agents 2 and 3.

  \item[(2)] Agent 1 sends more than one messages in round $r$. Suppose
    that agent 1 has successfully sent messages to agent 2 and 3 in
    round $r$.

By a similar argument as the previous case, we can prove that
	if we remove the
    message sent from agent 1 to agent 2 (or agent 3), the new dictator will
    become agent 2 (or agent 3). 
Now we remove the messages sent from agent 1 to
    both agent 2 and agent 3. 
Suppose in this case the new
    dictator is some agent $i$, and without loss of generality
	suppose that $i \neq 3$. 
Now we let
    agent 1 and agent 2 be in the colluding group. Then consider 
	the failure pattern $F'$ in which agent 1 has successfully sent 
	a message to
    agent 2 but not agent 3. 
Agent 2 can pretend that he did not
    receive the message from agent 1 in round $r$, 
	because he will know through the private channel at the end of 
	round $r$ that agent 1
    crashes at round $r$. 
In this case, the dictator will be
    agent $i$ instead of agent $3$, which could benefit agent $2$
	if $2$ prefers $i$'s most preferred value over agent $3$'s.
  \end{itemize}
\end{proof}

\section{Conclusion and future directions}
\label{sec:conclude}
In this paper, we propose new protocols that are resilient to both
	crash failures and strategic manipulations.
We argue that combining crash failures with
	strategic manipulations is an interesting research area addressing both
	practical scenarios and enriching the theory of fault-tolerant
	distributed computing.

There are many open problems and research directions one can look into.
First, with the problem 
	setting of this paper, several interesting open problems are left
	to be explored:
(a) whether a deterministic
	protocol resisting three or more colluders exist; and
(b) whether the gap between our round complexity ($2f+2$ or $3f+4$ depending
	on the cases) and the round complexity of standard protocols without
	self agents ($f+1$) can be closed, or there is an intrinsic cost
	in tolerating manipulations on top of crash failures.
Going beyond the setting of
	this paper, one can look into other utility functions such as
	message transmission costs, other distributed computing tasks, other
	distributed computing models such as asynchronous or shared memory
	systems, or other
	type of failures such as omission failures.
We wish that our work would stimulate other researchers to invest in
	this emergent area of incentive-compatible and fault-tolerant
	distributed computing.

\section*{Acknowledgement}

We are indebted to Shang-Hua Teng, whose insightful discussion with us 
	initiates our work on this topic.

\bibliographystyle{abbrv}
\bibliography{bibase}



\end{document}